\documentclass[useAMS,usenatbib]{mnras}
\usepackage{epsfig, color}
\usepackage{graphicx}
\usepackage[fleqn]{amsmath}
\usepackage{epstopdf}
\usepackage{caption}
\usepackage{float}

 \def\ltsima{$\;\buildrel < \over \sim \;$} 
 \def\simlt{\lower.5ex\hbox{\ltsima}}
 \def\gtsima{$\; \buildrel >\over\sim\;$}
 \def\simgt{\lower.5ex\hbox{\gtsima}}

\title[AGN jets on a moving mesh]{AGN jet feedback on a moving mesh:
  cocoon inflation, gas flows and turbulence} \author[]{Martin A.
  Bourne$^{\star}$ and Debora Sijacki \\ Institute of Astronomy and
  Kavli Institute for Cosmology, University of Cambridge, Madingley
  Road, Cambridge, CB3 0HA, UK\\ $^{\star}$ {E-mail:~} {\rm
    mabourne@ast.cam.ac.uk} }

\hypersetup{draft}

\begin{document} 
\date{Received} \pagerange{\pageref{firstpage}--\pageref{lastpage}}
\pubyear{2017} \maketitle \label{firstpage}

\maketitle

\begin{abstract} 
In many observed galaxy clusters, jets launched by the accretion process on to supermassive black holes, inflate large-scale cavities filled with energetic, relativistic plasma. This process is thought to be responsible for regulating cooling losses, thus moderating the inflow of gas on to the central galaxy, quenching further star formation and maintaining the galaxy in a red and dead state. In this paper, we implement a new jet feedback scheme into the moving mesh-code {\sc arepo}, contrast different jet injection techniques and demonstrate the validity of our implementation by comparing against simple analytical models. We find that jets can significantly affect the intracluster medium (ICM), offset the overcooling {through a number of heating mechanisms}, as well as drive turbulence, albeit within the jet lobes only. Jet-driven turbulence is, however, a largely ineffective heating source and is unlikely to dominate the ICM heating budget even if the jet lobes efficiently fill the cooling region, as it contains at most only a few percent of the total injected energy. We instead show that the ICM gas motions, generated by orbiting substructures, while inefficient at heating the ICM, drive large-scale turbulence and when combined with jet feedback, result in line-of-sight velocities and velocity dispersions consistent with the Hitomi observations of the Perseus cluster.
\end{abstract}


\section{Introduction}

Galaxy clusters show X-ray luminosities of up to $\sim 10^{45}$ erg s$^{-1}$ in their centres \citep{McNamara2007}, indicating significant energy losses. Based on early X-ray observations of galaxy clusters, it was estimated that the cooling time is shorter than the Hubble time.  At face value, therefore a cooling flow should be established with inflow rates of $\sim 1000$ M$_{\odot}$ yr$^{-1}$
\citep{Fabian94}.  Such cooling flows should result in significant
amounts of cold molecular gas and star formation within the central
cluster galaxies; however, this is not observed. Some mechanism (or
combination of mechanisms) is acting to prevent gas cooling and hence
maintain star formation rates at relatively modest values of $\sim
1$--$10$ M$_{\odot}$ yr$^{-1}$ \citep[e.g.][]{McNamara2007,
  DonahueEtAl10, CookeEtAl2016, FogartyEtAl17}, although in some
brightest cluster galaxies star formation can be $\simgt 100$
M$_\odot$ yr$^{-1}$ \citep[e.g.][]{CrawfordEtAl99,
  EgamiEtAl06, McNamaraEtAl06, vdLindenEtAl07, MittalEtAl15,
  FogartyEtAl17, MittalEtAl17}, with moderate molecular gas reservoirs within
cluster cores \citep[e.g.][]{ODeaEtAl07, McNamaraEtAl14,
  RussellEtAl14, RussellEtAl17}.  Furthermore, X-ray emission lines
below $1$ keV, expected for low temperature gas, were not seen in {\it ASCA}
\citep[e.g.][]{IkebeEtAl97, MakishimaEtAl01}, {\it XMM--Newton}
\citep[e.g.][]{PetersonEtAl01, PetersonEtAl03, TamuraEtAl01,
  BoehringerEtAl2002, MatsushitaEtAl02} or {\it Chandra}
\citep[e.g.][]{LewisEtAl2002} observations, which along with UV
observations \citep[e.g.][]{OegerleEtAl01, BregmanEtAl06} suggest
lower than expected cooling rates.

A number of mechanisms have been invoked to pump energy into the
intracluster medium (ICM) and explain the apparent lack of cooling,
including thermal conduction from hot gas at larger radii
\citep[e.g.][]{RuszkowskiBegelman02, ZakamskaNarayan03, VoigtFabian04,
  ConroyOstriker08, BogdanovicEtAl09, RuszkowskiOh10, RuszkowskiOh11}
and stirring by the motions of substructures \citep{FujitaEtAl04,
  RuszkowskiOh11}. However, the principle energy source is expected to
be feedback from an accreting supermassive black hole (SMBH). In
general, active galactic nucleus (AGN) feedback can be split into two
primary modes, both of which are thought to play an important role in
galaxy evolution \citep[see e.g.][for a review]{Fabian12}. The {\it
  quasar} mode, taking the form of powerful, isotropic winds and
outflows during phases of rapid BH growth, is believed to drive
observed BH scaling relations, such as the $M_{\rm BH}$--$\sigma$ and
$M_{\rm BH}$--$M_{\rm b}$ relations \citep[e.g.][]{Ferrarese00, Haering04, GultekinEtal09, KormendyHo13, McConnellMa13}, and
quench star formation. Whereas the {\it maintenance} mode, associated
with the production of jets by moderately accreting BHs, is thought to
keep the gas surrounding galaxies warm and hence prevent it from
cooling on to the galaxy. AGN feedback provides an explanation for the
observed discrepancy between the dark matter halo mass function and
the galaxy stellar mass function \cite[e.g.][]{BowerEtal06, Croton06}
and is invoked in analytical models and simulations to inhibit cooling
in galaxy clusters and ensure that early type elliptical galaxies
remain red and dead \citep[e.g.][]{BowerEtal06, Croton06,
  SijackiSpringel06, SijackiEtAl07}.

{X-ray} observations of galaxy clusters often show giant cavities of relativistic plasma \citep{FabianEtAl2000, FabianEtAl2011,
  McNamaraEtAl2000, HeinzEtAl2002, FormanEtAl07} that are expected to
be inflated by jets produced by accretion on to a central SMBH
\citep[e.g.][]{BinneyTabor95, OmmaEtAl04, McNamaraEtAl05,
  FabianEtAl06, SijackiSpringel06, CattaneoEtAl07,
  FormanEtAl07, SijackiEtAl07, DuboisEtAl10, DuboisEtAl12}. It is these jets and the
cocoons they inflate that are assumed to be the source of heating in
galaxy clusters \citep[e.g.][]{ChurazovEtAl01, ChurazovEtAl02,
  BirzanEtAl04}. The high fraction of cool core clusters that contain
cavities and exhibit radio emission \citep[see e.g.][]{Burns90,
  DunnEtAl05, DunnFabian06, DunnFabian08, McNamara2007, Sun09,
  Fabian12} suggest they are a common phenomenon.  The energy content
of these cavities ($10^{55}$--$10^{61}$ erg), based upon PV calculations, show a correlation with the X-ray luminosity/cooling
time within the ICM \citep{RaffertyEtAl06, McNamara2007, NulsenEtAl07,
  DunnFabian08, Fabian12, Hlavacek-LarrondoEtAl12} and are thus
expected to be the mechanism through which cooling is
regulated. Combining this with the fact that many clusters have short
central cooling times suggests that self-regulation is at play and
that feedback is the dominant mechanism regulating cooling and heating
\citep[see e.g.][for a full discussion]{McNamara2007}. On top of
this, many cool core clusters exhibit positive central temperature
gradients \citep[e.g.][]{CavagnoloEtAl08, HudsonEtAl10,
  McDonaldEtAl14}{,} further suggesting that any {\it central} heating
mechanism cannot exceed the rate of cooling.

{\it ROSAT} observations of the Perseus cluster provided the first clear
evidence for X-ray cavities \citep{BoehringerEtAl93}. Over the
following decades, {\it XMM--Newton} and {\it Chandra} delivered an ever growing
and more up to date collection of galaxy cluster observations
\citep[e.g.][]{FabianEtAl2000, FabianEtAl05, FabianEtAl06,
  McNamaraEtAl2000, TamuraEtAl01, HeinzEtAl02, FormanEtAl07,
  BlantonEtAl11, SandersEtAl2016}. Most recently, albeit short-lived,
the ill-fated Hitomi mission has provided the most detailed kinematic
observations of the Perseus cluster to date \citep{Hitomi2016},
showing a relatively sedate ICM, with respect to gas motions. Combined, these
observations have produced a wealth of information to aid our
understanding of how X-ray cavities are inflated and interact
with the ICM.

However, while such observations have provided valuable insight into
the processes at play, they contain limited temporal information on
the jet feedback process. Further, given observational difficulties in
observing faint cavities, such as those at large distances from the
cluster centre and in systems at high redshift, they provide an
inherently biased view of such systems.  With this in mind, it is vital
to couple the latest observational results with state-of-the-art
computer simulations.  Much numerical work has already been performed
in an attempt to understand various aspects of AGN jet feedback. The
large dynamic range involved in such simulations has, however, meant
that simplifications often have to be made.  Some simulations attempt
to mimic the inflation of jet cavities by injecting off centre, hot
bubbles \citep[e.g.][]{QuilisEtAl01, SijackiSpringel06,
  SijackiEtAl07, SijackiEtAl2015}, finding that this can effectively
disrupt cooling flows and reduce star formation rates, even in fully
self-consistent cosmological simulations. Other work, however, has
specifically included the inflation and evolution of the jet cavities
themselves and their subsequent impact on cluster haloes
\citep[e.g.][]{ChurazovEtAl01, ReynoldsEtAl01, BassonAlexander03,
  OmmaEtAl04, GaiblerEtAl09, DuboisEtAl11b, HardcastleKrause13,
  HardcastleKrause14, EnglishEtAl16, WeinbergerEtAl17}, focusing
mostly on isolated halo models. Further, simulations including
self-consistent feedback, in which the jet properties are linked to
estimated SMBH accretion rates \citep[e.g.][]{CattaneoEtAl07,
  DuboisEtAl10, DuboisEtAl12, GaspariEtAl11, GaspariEtAl12, LiBryan14, LiEtAl15,
  PrasadEtAl15, YangReynolds16a, YangReynolds16b}, find that
self-regulated SMBH growth and feedback is able to inhibit cooling and
produce a number of the observed properties of galaxy clusters.

Despite the successes of such models, there is still little consensus
on which processes dictate the transfer of the mechanical jet energy
isotropically to the ICM \citep[e.g.][]{VernaleoReynolds06} and hence
dominate the heating energy budget. A number of processes have been
suggested including heating due to dissipation of turbulence
\citep[e.g.][]{BanerjeeSharma2014, ZhuravlevaEtAl2014}, sound waves
\citep[e.g.][]{FabianEtAl03, FabianEtAl05b, FabianEtAl17,
  RuszkowskiEtAl04}, or gravity waves \citep[e.g.][]{OmmaEtAl04}, shock heating \citep[e.g.][]{FabianEtAl03,
  RandallEtAl15, LiEtAl2016}, mixing \citep[e.g.][]{HillelSoker16b,HillelSoker17},
cavity heating \citep[e.g.][]{ChurazovEtAl02, BirzanEtAl04}, cosmic
ray production \citep[e.g.][]{SijackiEtAl08, Pfrommer13}, and gas
circulation \citep[e.g.][]{YangReynolds16b}. However, as critically
pointed out by \citet{YangReynolds16b}, it is likely that a number of
different processes are at play. 

In this paper, we present a novel method of including AGN jet feedback
within the moving mesh-code {\sc arepo} \citep{SpringelArepo2010}. In Section
\ref{sec:numerical_method}, we discuss the {\sc arepo} code and outline our
new jet inflation prescription, while in Section \ref{sec:fiducial_runs}, we
present our first results, highlighting the robustness of our scheme
by carrying out a numerical resolution study and comparing different
jet injection techniques. In Section \ref{sec:prec_jets}, we consider jet
precession and discuss the properties of inflated jet cavities,
resultant gas flows and how halo properties are impacted. In Section
\ref{sec:sub_structure}, we include substructures in our galaxy cluster
{in order to} investigate the turbulence they drive
and how this impacts on jet evolution and properties. Specifically, in
this section, we compare recent Hitomi observations of the velocity
field in the Perseus cluster with results from our
simulations. Finally, in Section \ref{sec:discussion} and Section
\ref{sec:summary}, we discuss and summarize the results of our
simulations, highlighting implications for our understanding of the
role of jets in galaxy cluster evolution.

\section{Numerical method}
\label{sec:numerical_method} 
\subsection{Code}

We use the moving mesh-code {\sc arepo}, presented in
\citet{SpringelArepo2010}, which employs a finite-volume solver for
hydrodynamics and the {\sc treepm} method for gravity
\citep{Springel05}. The simulations presented in this paper rely on a
key ability of the hydrodynamic solver in {\sc arepo}, namely the
ability to refine and de-refine cells based on certain criteria and
thus allowing adequately high resolution where needed, but
computationally less expensive low resolution where possible. For most
of the simulations presented in this paper, we employ primordial
cooling and the subgrid interstellar medium (ISM)/star-formation model of
\citet{Springel03}, which implements an effective equation of state to
model the unresolved, supernova regulated, multiphase ISM and the
explicit (stochastic) formation of stars as {\it N-body} particles (in
actuality, each {\it N-body} particle represents a stellar population). We
note that, as in \cite{CurtisSijacki15}, gas within the central
refinement region (see Section \ref{sec:refinement}) is not allowed to form
stars in order to avoid spurious {\it N-body} heating effects. On top of
this, we implement (and modify) a number of other techniques important
to the BH growth and feedback, which we discuss in the
following sections. 

\subsection{Blackhole accretion and feedback}

All simulations presented below contain a central BH, modelled as a
sink particle, which can accrete surrounding gas, acts as a source of
feedback and is the BH refinement focal point (see Section
  \ref{sec:refinement}).

\subsubsection{Accretion}

The net growth rate of a BH can be calculated as the difference
between the gas accretion rate and mass outflow rate close to the BH
such that
\begin{equation}
\dot{M}_{\rm BH}=\left(1-\epsilon_{\rm r}\right)\dot{M}_{\rm
  a}-\dot{M}_{\rm J}
\label{mdot_bh_one}
\end{equation}
where $\epsilon_{\rm r}$ is the radiative efficiency of accretion,
$\dot{M}_{\rm a}$ is the gas accretion rate and $\dot{M}_{\rm J}$ is
the rate of mass outflow in the form of a jet.  Note that galaxy
formation and cosmological simulations typical neglect this final
term; however, as highlighted by \citet{OstrikerEtAl10a}, it can be
important to explicitly consider the mass outflow rate when
considering AGN feedback. Similarly to \citet{OstrikerEtAl10a}, we
define a jet mass loading factor
\begin{equation}
\eta_{\rm J} =\frac{\dot{M}_{\rm J}}{\dot{M}_{\rm BH}}.
\label{jet_mass_loading}
\end{equation}
Combining equations (\ref{mdot_bh_one}) and (\ref{jet_mass_loading})
results in a BH growth rate of
\begin{equation}
\dot{M}_{\rm BH} = \frac{1-\epsilon_{\rm r}}{1+\eta_{\rm
    J}}\dot{M}_{a}.
\label{mdot_bh_two}
\end{equation}
In what follows, BHs grow at the rate defined by equation
(\ref{mdot_bh_two}) where we set $\eta_{\rm J}=1$,
$\epsilon_{\rm r}=0.1$ and assume that $\dot{M}_{\rm a}$
is a fixed fraction of the Eddington accretion rate
\begin{equation}
\dot{M}_{\rm Edd}=\frac{4\pi GM_{\rm BH}m_{\rm p}}{\epsilon_{\rm
    r}\sigma_{\rm T} c},
\end{equation}
where $G$ is the gravitational constant, $m_{\rm p}$ is the proton
rest mass, $\sigma_{\rm T}$ is the Thompson scattering cross-section
and $c$ is the speed of light. As such, in this work, we fix
$\dot{M}_{a}=0.02\dot{M}_{\rm Edd}$, and so from equation
(\ref{mdot_bh_two}) $\dot{M}_{\rm BH}=0.009\dot{M}_{\rm Edd}$,
corresponding to a jet power of $\sim 10^{45}$ erg
s$^{-1}$ for $M_{\rm BH}=10^{9}$ M$_{\odot}$. While we do
not consider self-consistent BH growth and feedback here, we will
investigate self-regulation in future work. 

\subsubsection{Physical properties of the jet energy and momentum content}

Physically, accretion on to the BH is expected to release
energy at a rate of
\begin{equation}
\dot{E}_{\rm J}=\epsilon_{\rm f}\epsilon_{\rm r}\dot{M}_{\rm
  BH}c^{2}\left(=\frac{1}{2}\dot{M}_{\rm J}v_{\rm J}^{2}\right),
\end{equation}
where $\epsilon_{\rm f}=1$ is the efficiency of coupling
energy to the jet and $\dot{M}_{\rm J}v_{\rm J}^{2}/2$ is the jet
kinetic energy. Combining equations (\ref{jet_mass_loading}) and
(\ref{mdot_bh_two}), the mass outflow rate in the jet can be written
as
\begin{equation}
\dot{M}_{\rm J}=\eta_{\rm J}\frac{1-\epsilon_{\rm r}}{1+\eta_{\rm
    J}}\dot{M}_{a},
\label{eq:jet_mass_rate}
\end{equation}
while the kinetic energy and momentum of the jet are given as
\begin{equation}
\dot{E}_{\rm J}=\frac{1-\epsilon_{\rm r}}{1+\eta_{\rm J}}\epsilon_{\rm
  f}\epsilon_{\rm r}\dot{M}_{\rm a}c^{2},
\label{eq:jet_energy_rate}
\end{equation}
and
\begin{equation}
\dot{p}_{\rm J}=\eta_{\rm J}\frac{1-\epsilon_{\rm r}}{1+\eta_{\rm
    J}}\left(\frac{2\epsilon_{\rm f}\epsilon_{\rm r}}{\eta_{\rm
    J}}\right)^{1/2}\dot{M}_{\rm a}c,
\label{eq:jet_momentum_rate}
\end{equation}
respectively, where the velocity of {the} sub-resolution
  jet would be
\begin{equation}
{v_{\rm J}}=\left(\frac{2\epsilon_{\rm f}\epsilon_{\rm r}}{\eta_{\rm
    J}}\right)^{1/2}c\simeq 0.447 c \left(\frac{\epsilon_{\rm
    f}}{1}\right)^{1/2}\left(\frac{\epsilon_{\rm
    r}}{0.1}\right)^{1/2}\left(\frac{\eta_{\rm J}}{1}\right)^{-1/2}.
\label{eq:jet_vel}
\end{equation}
Note that given we do not resolve the jet on very small scales, velocities never exceed $\sim 0.25c$ in our simulations. As
  such we expect that relativistic dynamical effects would make a negligible difference.

\subsubsection{Simulated jet properties and structure}
\label{sec:sim_jet_props}

There are a number of methods in the literature used for injecting jet
energy on scales resolved in galaxy cluster
  simulations. Broadly speaking they either only inject kinetic
energy \citep[e.g.][]{DuboisEtAl10, GaspariEtAl11, YangReynolds16a,
  YangReynolds16b}, or some combination of momentum/kinetic energy
plus thermal energy \citep[e.g.][]{CattaneoEtAl07, LiBryan14}. Each has advantages and drawbacks, which we discuss in Section \ref{sec:discussion_numerical_jets}. {We also note that jets have been simulated in other astrophysical scenarios, such as in star formation \citep[e.g.][]{FederrathEtAl14}, that use similar injection techniques.} As such, in this paper we consider three main types of jet energy injection, which we term as
momentum, thermal and kinetic jets for when only momentum,
momentum plus thermal energy {or} purely kinetic energy {is injected},
respectively. On resolved scales, {the jet} is injected into a
cylinder  centred on the BH, with a variable radius, $r_{\rm Jet}$. This
  radius is varied such that the cylinder contains a
fixed {target} gas mass, $M_{\rm Jet}$, which for all simulations presented here
is set to $M_{\rm Jet}=10^{4}$ M$_{\odot}$, although we present the
impact of other jet masses in Appendix \ref{App:JetMass}. The
cylinder is  divided into two halves (north and south), each with radius, $r_{\rm Jet}$ and height, $h_{\rm
  Jet}$ (such that the total cylinder length is $2h_{\rm Jet}$), which are defined to have a fixed ratio $r_{\rm
  Jet}/h_{\rm Jet}=\tan\left(\theta_{\rm Jet}/2\right)$, where
$\theta_{\rm Jet}$ is the jet opening angle. For simulations presented
here, $r_{\rm Jet}/h_{\rm Jet}=3/2$, such that the total cylinder
volume is $\left(4/3\right)\pi r_{\rm Jet}^{3}$. For all three energy
injection regimes, half of the jet material is injected into the cells within each half of the
cylinder, weighted according to a kernel function similar to that
already used in the literature \citep[e.g.][]{OmmaEtAl04,
  CattaneoEtAl07, YangEtAl2012} of the form\footnote{We discuss an
  alternative kernel weighting scheme and its impact on jet evolution
  in Appendix \ref{App:Kernel}.}
\begin{equation}
W_{\rm J}(r, z)\propto{\exp\left(-\frac{r^{2}}{2r_{\rm
      Jet}^{2}}\right)|z|}.
\label{eq:kernel}
\end{equation}
The mass injected into an individual cell, $i$, is given as
\begin{equation}
{\rm d}m_{i}=\frac{\dot{M}_{\rm J}{\rm d}t}{2}\frac{m_{\rm i}W_{\rm J}(r, z)}{M_{\rm
    Weight}},
\end{equation}
where $\dot{M}_{\rm J}$ is given by equation (\ref{eq:jet_mass_rate}),
d$t$ is the BH timestep, $m_{i}$ is the cell mass, $M_{\rm
  Weight}=\sum_{i}{m_{i}W_{\rm J}(r, z)}$ is the weighted sum of cell masses in the relevant half-cylinder and the factor $1/2$ is to account for injecting half of the jet material into each half of the cylinder. For both the momentum and thermal
jets, momentum is added to cells within the cylinder following the
same weighting as the mass, such that the change in momentum of an
individual cell is given as
\begin{equation}
{\rm d{\textbf{\textit{p}}}_{i}}=\frac{\dot{{\textbf{\textit{p}}}}_{\rm J}{\rm d}t}{2}\frac{m_{\rm i}W_{\rm
    J}(r, z)}{M_{\rm Weight}},
\label{eq:inj_cell_momentum}
\end{equation}
where $\dot{{\textbf{\textit{p}}}}_{\rm J}$ is given by equation
(\ref{eq:jet_momentum_rate}). The injection of mass and momentum
results in a change in the kinetic energy of a cell of
\begin{equation}
    {\rm d}E_{i}^{kin}=\frac{({\textbf{\textit{p}}}_{i,0}+
        {\rm d}{\textbf{\textit{p}}}_{i})^{2}}{2(m_{i,0}+{\rm d}m_{i})} - \frac{
        {\textbf{\textit{p}}}_{i,0}^{2}}{2m_{i,0}}.
 \end{equation}
For the thermal and kinetic jets, we also calculate the total expected
energy injected into each cell as
\begin{equation}
{\rm d}E_{i}^{tot}=\frac{\dot{E}_{\rm J}{\rm d}t}{2}\frac{m_{\rm i}W_{\rm J}(r, z)}{M_{\rm
    Weight}},
\end{equation}
where $\dot{E}_{\rm J}$ is given by equation
(\ref{eq:jet_energy_rate}). Due to mass loading and momentum
cancellation ${\rm d}E_{i}^{kin} < {\rm d}E_{i}^{tot}$ and thus we correct for
this in the thermal and kinetic models to ensure energy
conservation. In the case of the thermal jet, we inject internal
energy equal to ${\rm d}E_{i}^{therm}={\rm d}E_{i}^{tot}-{\rm d}E_{i}^{kin}$. In the
case of the kinetic jet, instead of injecting momentum given by
equation (\ref{eq:inj_cell_momentum}), for each cell we calculate a
momentum kick of magnitude
\begin{equation}
|{\rm d}{\textbf{\textit{p}}}_{i}|=\sqrt{2(m_{i,0}+dm_{i})(E_{i,0}+dE_{i}^{tot})}-|{\textbf{\textit{p}}}_{i,0}|,
\end{equation}
which is added to the current momentum of the cell, along the
direction of the jet. A subtlety to note here is that the final
momentum of the cell will only equal
$\sqrt{2(m_{i,0}+{\rm d}m_{i})(E_{i,0}+{\rm d}E_{i}^{tot})}$, if the initial cell
momentum vector and the jet momentum vector are aligned; otherwise,
momentum cancellation will still result in a small loss of kinetic
energy. We correct for any such loss by injecting additional thermal
energy to ensure total energy conservation.

\begin{figure}
\psfig{file=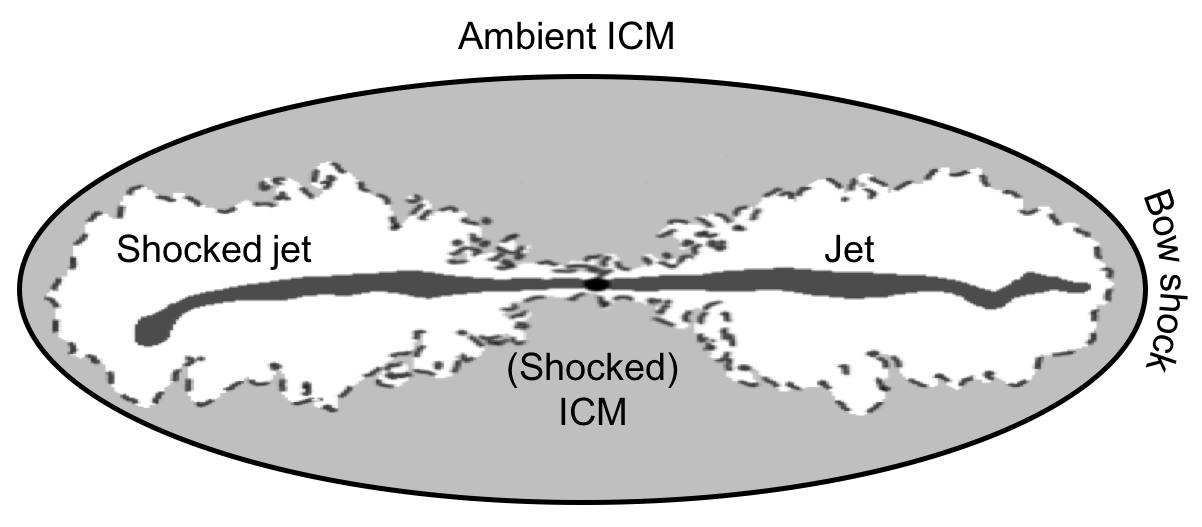, width=0.45\textwidth,angle=0}
\caption{Schematic of the main jet components considered in this
  study, including the jet lobes, which consist of the jet and shocked
  jet material, the (shocked) ICM region, which completes the jet cocoon, and
  the surrounding ambient ICM. }
\label{fig:jet_structure}
\end{figure}

The vicinity of the jet can be split into four main regions, as
illustrated in Fig. \ref{fig:jet_structure}, which provides a
schematic of the typical jet environment. We will refer throughout
this paper to the regions as the jet, shocked jet material, (shocked)
ICM and ambient ICM. {The first three of these make up} the jet cocoon. During all simulations, jet material is tracked
using a tracer field, where each cell tracks the jet material mass,
$m_{\rm J}$, within the cell and a jet mass fraction parameter,
$f_{\rm J} = m_{\rm J}/m_{\rm cell}$, where $m_{\rm cell}$ is the
total gas mass within the cell. $m_{\rm J}$ is set to $m_{\rm cell}$
for any cell into which jet material, momentum and energy is directly
injected {(i.e., cells within the jet cylinder)}, with $m_{\rm J}$ being advected in line with the total gas
mass. When considering the properties of the jet, similar to
\citet{YangReynolds16a} we define the jet lobe material as those cells
with $f_{\rm J} > 0.01$ and the jet length as $[{\rm max}(z_{\rm
    J})-{\rm min}(z_{\rm J})]/2$, where $z_{\rm J}$ are the $z$ co-ordinates
of jet material. We note that other works in the literature
\citep[e.g.][]{HardcastleKrause13, WeinbergerEtAl17} use a threshold
of $f_{\rm J} > 0.001$. While this can impact the inferred energy
content of the jet lobes (see Section \ref{sec:energy_mom_comp}), it has a
negligible effect on the measured jet length, which is rather
insensitive to the chosen value of $f_{\rm J}$.

\subsubsection{Super-Lagrangian refinement technique}
\label{sec:refinement}

The simulations presented in this study rely heavily upon the recently
published Super-Lagrangian refinement (SLR) technique of
\citet{CurtisSijacki15}. The method allows grid cells to be refined in
the vicinity of a BH according to predefined criteria for cell sizes
as a function of distance from the BH. This results in significantly
improved resolution close to the BH allowing for more accurate
estimates of the accretion rate on to the BH and, as shown in this
paper, the ability to model AGN feedback in the form of jets. As in
\citet{CurtisSijacki15}, the SLR region has an outer radius $h_{\rm
  BH}$, which is defined as the region surrounding the BH that
contains a total mass of gas cells of $n_{\rm ngb}^{\rm BH}\times
m^{\rm target}_{\rm cell}$, where $n_{\rm ngb}^{\rm BH}$ is the number
of neighbouring gas cells to the BH (without SLR) and is set to $32$ in
all runs except for resolution tests, and $m^{\rm target}_{\rm cell}$
is the target cell mass of a given simulation. In order to allow for
SLR of cells around the BH to take effect, jets are only activated
after $\sim 2.45$ Myr in all simulations. 

Despite the SLR scheme, we found that when high-density gas flows into
the vicinity of the jet, gas cells can have masses comparable to (or
in excess) of $M_{\rm Jet}$. Therefore, to reduce the risk
  of under populating the jet cylinder we have implemented an
additional jet refinement (JR) scheme during the jet injection
process. Therefore at a radius of $\gamma r_{\rm Jet}$, the maximum
allowed cell mass, $m_{\rm cell}^{\rm max}$, is equal to the target jet
cylinder mass, $M_{\rm Jet}$, and decreases {for} smaller radii following
the power law
\begin{equation}
  \frac{m_{\rm cell}^{\rm max}}{M_{\rm Jet}}=(\alpha -
  \beta)\left(\frac{r}{r_{\rm Jet}}\right)^{\kappa} + \beta,
\end{equation}
where
\begin{equation}
  \kappa = \ln\left(\frac{1-\beta}{\alpha-\beta}\right)-\ln\gamma,
\end{equation}
such that $m_{\rm cell}^{\rm max}=\alpha M_{\rm Jet}$ at $r=r_{\rm
  Jet}$ and $m_{\rm cell}^{\rm max}=\beta M_{\rm Jet}$ at $r=0$. Most
of our runs use $\alpha = 0.01$, $\beta = 0.001$ and $\gamma = 3$,
which results in the jet cylinder typically being populated by $\sim 200$
cells. Note that in the remainder of this paper, all simulations
include this additional JR scheme while the jet is
active. 

\subsubsection{Cell draining}

During the course of simulations, mass is added both to the central BH
as it accretes material and into the jet cylinder. In order to
conserve mass, material is simultaneously removed from gas cells
outside of the jet cylinder, but within $r < h_{\rm BH}$, which lie
within a torus-shaped region with an opening angle $\left(\pi -
\theta_{\rm Jet}\right)$, in the plane perpendicular to the jet
axis. The exact method used to drain mass is presented in
\citet{CurtisSijacki15}, while for the generic draining procedure, see
\citet{VogelsbergerEtAl13}. In the simulations presented here, the
total mass drained per timestep is given by $M_{\rm drain} =
(\dot{M}_{\rm BH} + \dot{M}_{\rm J}){\rm d}t$, where $\dot{M}_{\rm BH}$ and
$\dot{M}_{\rm J}$ are given by equations (\ref{mdot_bh_two}) and
(\ref{eq:jet_mass_rate}), respectively. The necessary mass is removed
from eligible cells, with each cell contributing a mass of $M_{\rm
  drain}\times m_{\rm cell}/M_{\rm tot}$, where $M_{\rm tot}$ is the
total mass of eligible cells. An additional condition is imposed that
cells can only contribute {up to} $90$ per cent of their total mass. 

\section{Numerical implementation and verification}
\label{sec:fiducial_runs}

\subsection{Simulations and set-up}
 \label{sec:sim_setup}

 In this section we perform simulations to compare different numerical
 parameters and jet injection techniques. Apart from the simulations
 in Section \ref{sec:analytical}, all other simulations are performed
 within a static background potential that follows a
 \citet{Hernquist90} profile and an accompanying gas distribution that
 follows the same profile, except for a slightly softened
   core \citep[see e.g.][]{SijackiEtAl06a, SijackiEtAl07}. Save for
   the softened core, the total enclosed mass of the system initially
 follows:
 \begin{equation}
 M(r)=M_{200}\frac{r^{2}}{(r+a)^{2}},
 \label{eq:hernquist}
 \end{equation}
 where $M_{200}=10^{14}$ M$_{\odot}$, $a=175.98$ kpc and the gas mass
 makes up $f_{\rm g}=0.18$ of the total halo mass.  Except for
 resolution testing, the gas component, which extends out
   to $r=100a$, is modelled using $10^6$ cells, each with a target
 mass of $m^{\rm target}_{\rm cell}=1.8\times 10^{7}$ M$_{\odot}$. The
 system is initially set-up in hydrostatic equilibrium and relaxed
 non-radiatively for $\sim 5$ Gyr to produce the initial
 conditions. A BH particle of mass $M_{\rm BH}=10^{9}$ M$_{\odot}$ is
 then added to the centre of each halo to act as the source of the SLR
 and JR schemes, and as the source of jet feedback.
 
\begin{figure}
  \psfig{file=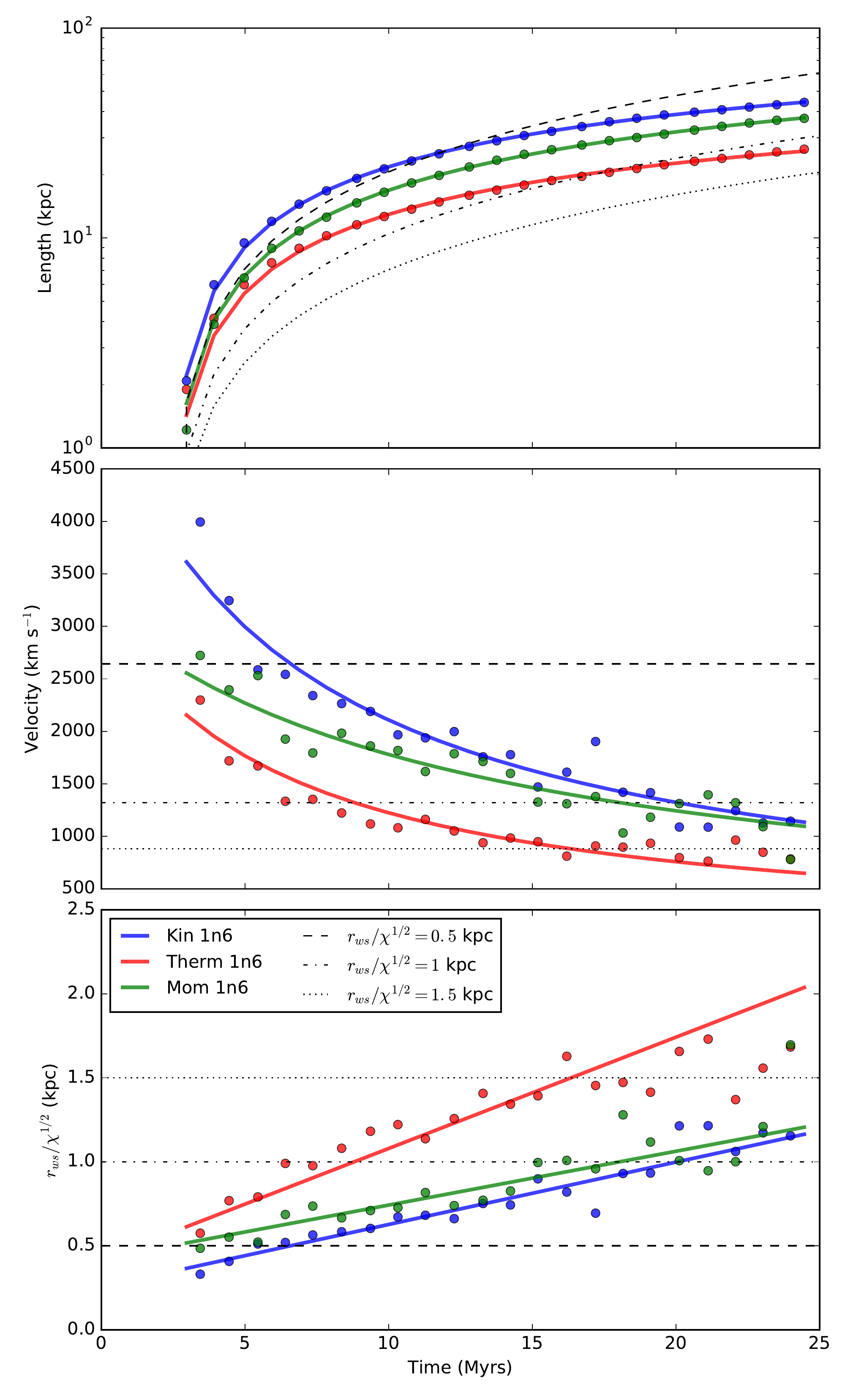,
    width=0.5\textwidth,angle=0}
  \caption{Overview: comparison of the evolution of simulated
    jet lengths and velocities to analytical solutions derived from
    the models of \citet{Begelman89}. Simulated jets differ slightly
    from this simple model because the working surface radius of the
    jet increases with time, resulting in a decelerating cocoon
    expansion. Top panel: evolution of jet length. Black curves
    show solutions to equation (\ref{eq:an_length}), assuming $r_{\rm
      ws}/\chi^{1/2}$ is fixed with values of $0.5$ (dashed), $1$
    (dot--dashed) and $1.5$ (dotted) kpc. Filled circles are measured
    directly from idealized simulations of a kinetic (blue), thermal
    (red) and momentum (green) jet, using the definition of jet length
    given in Section \ref{sec:sim_jet_props}, while the solid coloured
    lines show fits to the points, using equation
    (\ref{eq:fit_length}). Middle panel: evolution of vertical
    cocoon expansion velocity. Black lines show velocities calculated
    using equation (\ref{eq:an_vel}) corresponding to same $r_{\rm
      ws}/\chi^{1/2}$ values used in the top panel. The solid points
    show velocities calculated using the change in jet length between
    snapshots from the appropriate simulations while the solid--coloured lines show velocities calculated from the fits to jet
    length, using equation (\ref{eq:fit_vel}). Bottom panel:
    evolution of the estimated jet working surface
    radius, $r_{\rm ws}/\chi^{1/2}$. Black lines illustrate the values
    chosen for the simple analytical models. Solid points and coloured
    lines are estimates of $r_{\rm ws}/\chi^{1/2}$ from the length and
    velocity evolution deduced from simulations. They are calculated
    by plugging the velocities from corresponding points and lines in
    the middle panel into equation (\ref{eq:fit_r}). }
\label{fig:analytical}
\end{figure}
 
\subsection{Jet propagation: comparison to analytical models}
\label{sec:analytical}

Before delving into comparisons of different numerical parameters, we
first consider the expected propagation of jets from analytical
considerations. Analytical estimates for the evolution of jet
properties have previously been outlined, for example in
\citet{Begelman89}, who show that the $z$-component of the velocity of
the jet cavity can be estimated by balancing the thrust of the jet,
$\dot{M}_{\rm J}v_{\rm J}$, where $\dot{M}_{\rm J}$ and $v_{\rm J}$
are given by equations (\ref{eq:jet_mass_rate}) and
(\ref{eq:jet_vel}), respectively, with the force due to the ram
pressure of the ambient medium $\rho_{\rm ICM}v_{\rm c}^{2}\pi r_{\rm
  ws}^{2}$, where $v_{\rm c}$ is the velocity with which the cocoon
expands in the jet direction and $r_{\rm ws}$ is the working surface
radius of the jet. Given that $\dot{p}_{\rm J}=\dot{M}_{\rm J}v_{\rm
  J}$, the large-scale jet velocity (for a single jet) can be calculated as
\begin{equation}
v_{\rm c}= \left(\frac{\chi\dot{p}_{\rm J}}{2\rho_{\rm
    ICM}\pi}\right)^{1/2}\frac{1}{r_{\rm ws}},
\label{eq:an_vel}
\end{equation}
where $\chi$ is a momentum boost factor that may arise, for example,
due to mass loading of the jet. Equation (\ref{eq:an_vel}) is
straightforward to solve if we assume that all of the variables are
constant, giving a solution for the jet length at time $t$ of
\begin{equation}
 l_{\rm c}=\left(\frac{\chi\dot{p}_{\rm J}}{2\rho_{\rm
     ICM}\pi}\right)^{1/2}\frac{t-t_{0}}{r_{\rm ws}} + l_{0},
 \label{eq:an_length}
 \end{equation}
 where {$t_{0}$ is the jet start time and }$l_{0}$ is the initial jet length, which we set to the initial
 height of the jet cylinder, i.e. $262$ pc. This solution is plotted
 in the top panel of Fig. \ref{fig:analytical} by the black curves
 for jets with $r_{\rm ws}/\chi^{1/2}=0.5$ (dashed), $1$ (dot--dashed)
 and $1.5$ kpc (dotted). Also plotted with the filled circles in the
 top panel are average jet lengths  measured from
 simulations for kinetic (blue), thermal (red) and momentum (green)
 jets, using a simplified set of simulations compared to our other
 runs. Here, we model the ICM of a constant density and do not include
 a background potential or additional physics such as star formation
 or radiative cooling. This allows {a} more meaningful interpretation when
 comparing to the analytical model. While the simulated jet lengths sit within a sensible range of values when compared to the analytical solutions, none of them follow a single analytical track. To understand this further, we consider the evolution of the jet
 velocity, which for constant $r_{\rm ws}/\chi^{1/2}$ is also
 constant. However, the evolution of jet length found in the
 simulations is well fit if we assume $r_{\rm ws}/\chi^{1/2}$ scales
 linearly with time such that the jet velocity is of the form
\begin{equation}
v_{\rm c}= \left(\frac{\dot{p}_{\rm J}}{2\rho_{\rm
    ICM}\pi}\right)^{1/2}\frac{1}{at+b},
\label{eq:fit_vel}
\end{equation}
where $a$ and $b$ are parameters describing the evolution of the jet
working surface radius and momentum boost, $r_{\rm ws}/\chi^{1/2}$,
and hence the length of the jet evolves as
\begin{equation}
 l_{\rm c}=\left(\frac{\dot{p}_{\rm J}}{2\rho_{\rm
     ICM}\pi}\right)^{1/2}\frac{1}{a}\ln\left(\frac{at+b}{at_{0}+b}\right)
 + l_{0}.
 \label{eq:fit_length}
 \end{equation}
We fit the jet length evolution from the simulation with equation
(\ref{eq:fit_length}), with $a$ and $b$ set as free parameters {and assume that $\dot{p}_{\rm J}$ is constant}\footnote{Strictly speaking the BH accretes mass during the simulation resulting in a slight increase in $\dot{p}_{\rm J}$ over time; however, this is negligibly small ($<0.5$ per cent).}. The fits are shown by the solid lines in the top panel of Fig. \ref{fig:analytical}. The middle panel then shows the evolution
of the jet velocity; the solid lines are from equation
(\ref{eq:fit_vel}) using $a$ and $b$ calculated from fitting the jet
length evolution, while the filled points are velocities calculated
using the change in jet length between consecutive snapshots. The
horizontal black lines illustrate the analytic jet velocities for
$r_{\rm ws}/\chi^{1/2}=0.5$ (dashed), $1$ (dot--dashed) or $1.5$ kpc
(dotted). This panel directly shows that the simulated jet velocity
generally decreases with time. The changes in velocity can be
attributed to an evolution of $r_{\rm ws}/\chi^{1/2}$ with time. 

From the velocities, one can estimate the radius of the jet working
surface (with $\chi$ dependence) as 
\begin{equation}
  \frac{r_{\rm ws}}{\chi^{1/2}} =  \left(\frac{\dot{p}_{\rm
      J}}{2\rho_{\rm ICM}\pi}\right)^{1/2}\frac{1}{v_{\rm c}}.
  \label{eq:fit_r}
\end{equation}
We have estimated the working surface radii for
velocities measured from both the fits to the jet length evolution and {from}
velocities measured between snapshots; these are plotted in the lower
panel of Fig. \ref{fig:analytical} with the solid lines and coloured
circles, respectively. $\chi=1$ for both the thermal and
momentum-driven jets, while we estimate that it varies between $2$ and $3$
for the kinetic jet.  Therefore, the bottom panel
suggests that the working surface radius of the jet
increases with time. We note that while the jet cylinder radius in these simulations generally increases with time (due to the central densities falling), we also expect the jet to broaden naturally as it propagates to larger distances, a result which has also been found by previous
work \citep[e.g.][]{NormanEtAl82, LindEtAl89, KrauseCamenzind01,
  Krause03} and is the likely reason why the
evolution of jet length found in our simulations differs somewhat from
that predicted by simple analytical arguments
\citep[e.g.][]{Begelman89}.

\subsection{Jet parameters -- a numerical study}

\subsubsection{Different refinement schemes}
\label{sec:deref}

\begin{figure*}
\psfig{file=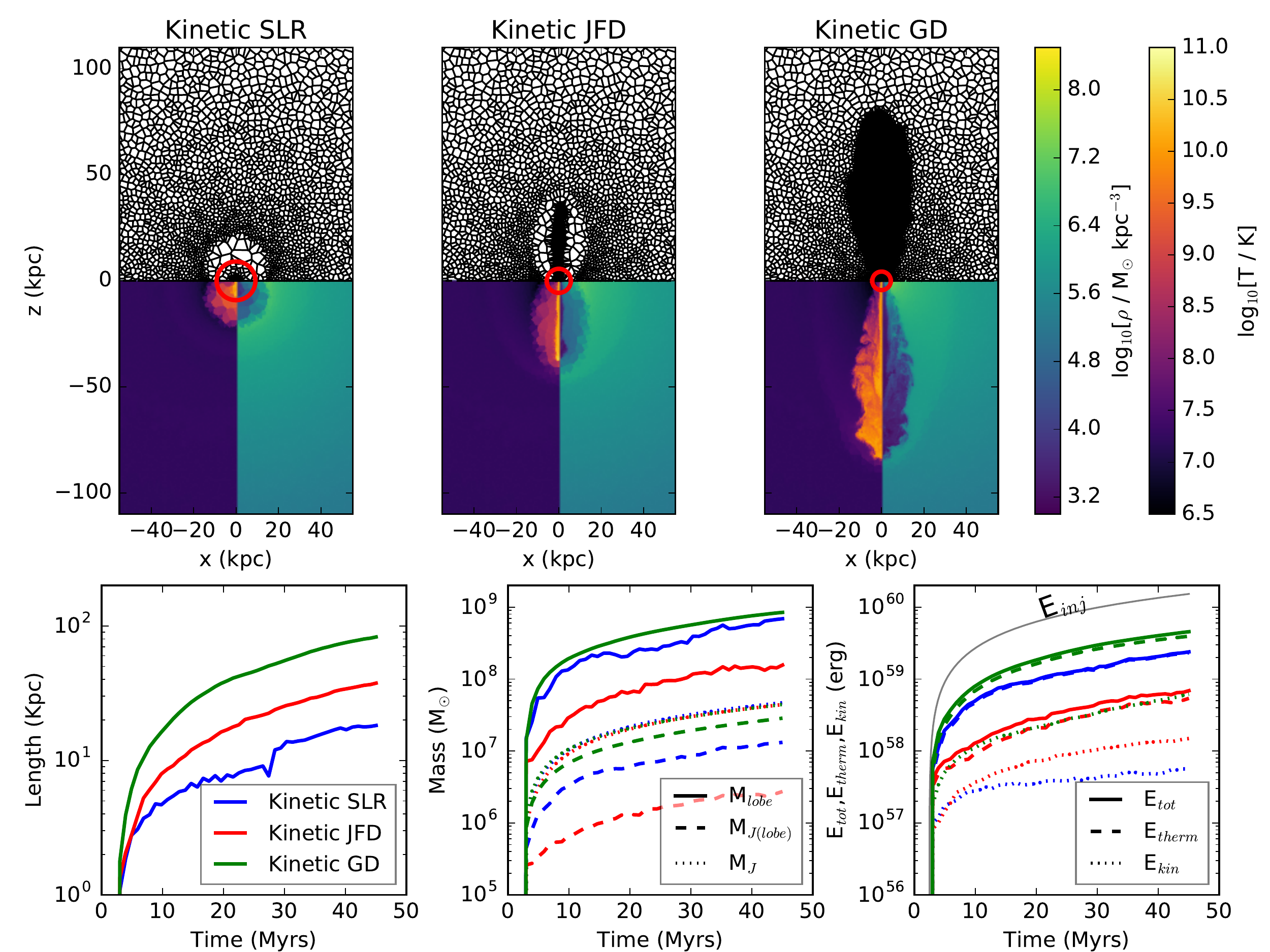,
  width=1.\textwidth,angle=0}
\caption{Overview: the impact of different refinement schemes on
  jet evolution. Reducing the occurrence of de-refinement allows the
  propagation of a jet with well-resolved interactions with the
  ICM. Top row: the top half of the panels show reconstructed
  schematics of Voronoi cell structure in the $y=0$ plane for kinetic
  jet runs with standard de-refinement (SLR), no de-refinement for jet
  material [$f_{\rm jet} > 0.01$, jet fraction de-refinement (JFD)] and gentle de-refinement (GD)
  from left to right respectively, at $t\simeq 45$ Myr. Note that the
  BH refinement region is indicated by the red circle, which has a
  radius of $h_{\rm BH}$. The bottom half of the top panels show
  density and temperature slices through the $y=0$ plane at $t\simeq
  45$ Myr for the respective runs. Bottom row: evolution of the
  jet length (left-hand panel), different components of jet mass
  (middle panel) and different components of jet energy content
  (right-hand panel) for the SLR (blue), JFD (red) and GD (green)
  jets. For comparison, we show the total injected jet energy
  (equation \ref{eq:e_inj}) by the solid black line in the lower
  right-hand panel.}
\label{fig:jet_overview_deref}
\end{figure*}

When performing astrophysical simulations, there is unfortunately not
a {\it one size fits all} refinement scheme and as such one has to
devise a specific scheme appropriate for the problem at hand. Here, we
consider the refinement and de-refinement criteria necessary to ensure
that we can resolve the jet to a suitable level and ensure it is able
to propagate to large scales. Without any additional
  refinement, the typical cell mass of $\sim 1.8\times 10^{7}$
M$_{\odot}$ is considerably larger than $M_{\rm Jet}$. The SLR
technique (in addition to the added JR scheme) therefore
allows us to inject the jet into a much smaller mass and hence not
excessively dilute the jet properties.  However, additional steps need
to be taken in addition to the SLR technique outlined in Section
\ref{sec:refinement}. To aid the discussion of these modifications, we
refer to Fig. \ref{fig:jet_overview_deref}, which shows in the top
half of the top panels, a reconstructed schematic of the $2$D Voronoi
cell structure in the $y=0$ plane for a kinetic jet run with various
de-refinement techniques (see discussion below).  The figure is
produced by considering the cells that exist in the $y=0$ plane and
then constructing a $2$D Voronoi grid based upon the mesh generating
points of those cells.  The left-hand panel shows the pure SLR
(+JR) model outlined in Section \ref{sec:refinement}. Here the level of refinement gradually reduces for cells further away from
the BH until they are beyond the SLR region, defined by the smoothing
length of the BH. At this point, the cells (de-)refine based upon a
mass criteria. Unmodified, this results in the jet cells produced at
small radii being de-refined, as they propagate away from the BH. This
is illustrated in the left-hand panel of Fig. \ref{fig:jet_overview_deref}, in which the cells become larger with
increased distance from the BH. Indeed, cells just beyond $h_{\rm
  BH}$, indicated by the red circles, that are de-refined such that
$m_{\rm cell}\simeq m_{\rm target}$, are somewhat larger than other
ambient gas cells, due to their increased temperature and reduced
density. Importantly, no visible jet is produced with this refinement
scheme. The central panel of Fig. \ref{fig:jet_overview_deref}
illustrates a second de-refinement scheme in which we modify the SLR
scheme such that cells can only de-refine if $f_{J} < 0.01$, which we
will refer to as the jet fraction de-refinement (JFD) scheme.  In this
case, a column of high resolution cells can be seen along the $z$-axis,
which are {\it jet} cells with $f_{J} > 0.01$; however, beyond these
the de-refinement acts in the same way as in the SLR scheme
illustrated in the left-hand panel. Similarly, there is a population of
large, low-density and high temperature cells that are (de-)refined
based on the mass criteria. Finally, the right-hand panel of Fig.
\ref{fig:jet_overview_deref} illustrates the grid structure when we
employ a gentle de-refinement scheme in addition to the JFD scheme,
whereby cells can only merge if gradients between neighbouring cells
are suitably small. We refer to this as the gentle de-refinement (GD)
scheme. This results in a jet consisting purely of high-resolution
cells that can propagate to large distances and in which
instabilities are not washed out. We note that as the occurrence of
the de-refinement is reduced $h_{\rm BH}$ becomes smaller. We
attribute this to the heating becoming less isotropic, and thus, there
is an increased inflow of gas to the cluster centre perpendicular to
the jet direction, and hence, the central density is increased.

\begin{figure*}
\psfig{file=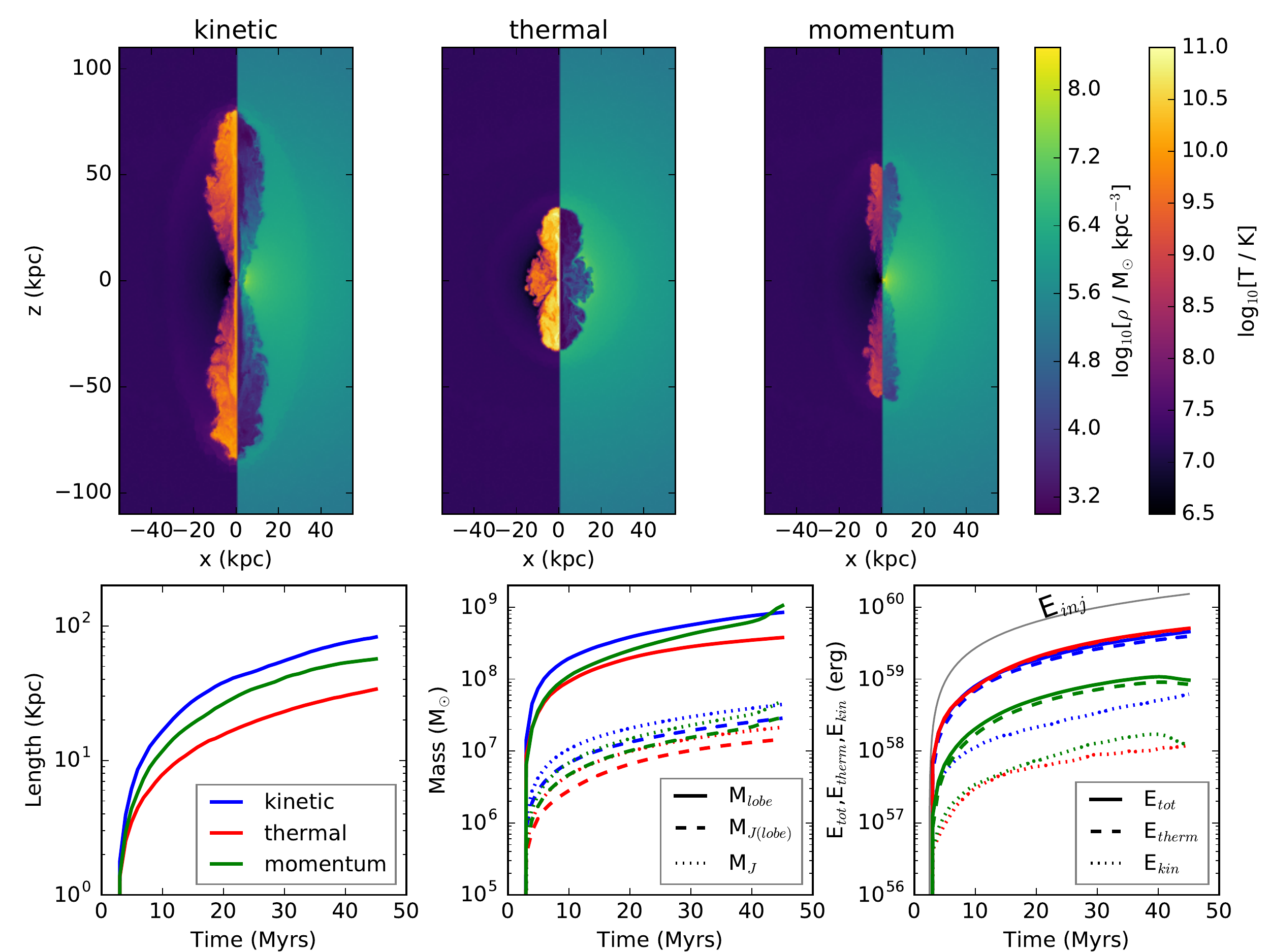,
  width=1.\textwidth,angle=0}
\caption{Overview: dependence of jet evolution on energy
  injection mechanism. While different mechanisms result in different
  morphological properties of the jet, the energy contents remain
  similar. Top row: density and temperature slices through the
  $y=0$ plane at $t\simeq 45$ Myr. Bottom row: evolution of the
  jet length (left-hand panel), different components of jet mass
  (middle panel) and different components of jet energy
  content(right-hand panel) for the kinetic (left-hand panel and blue
  curves), thermal (middle panel and red curves) and momentum
  (right-hand panel and green curves) runs. For comparison, we show
  the total injected jet energy (equation \ref{eq:e_inj}) by the solid
  black line in the lower right-hand panel.}
\label{fig:jet_overview_norm}
\end{figure*}

\begin{figure}
\psfig{file=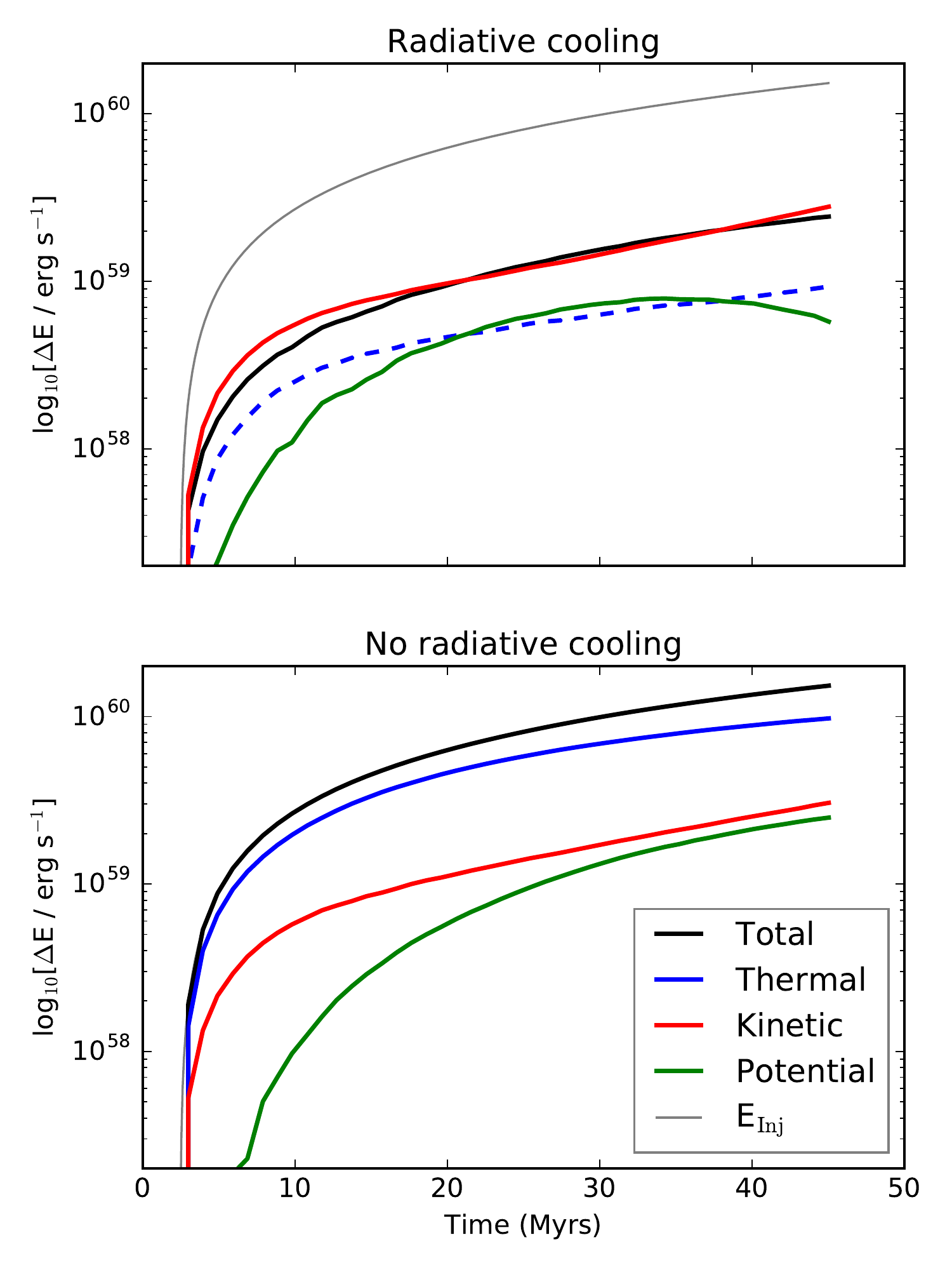,
  width=0.49\textwidth,angle=0}
\caption{Overview: change in energy content within a sphere with gas mass equal to the mass of gas initially within the central
  $125$ kpc of the galaxy cluster, during the injection of a kinetic
  jet. Solid lines show energy gains while dashed lines show energy
  losses. {\it Top panel:} run with radiative cooling. The ICM gas
  loses thermal energy due to radiative cooling (blue dashes
  curve). However the rate of radiative cooling is reduced by the jet
  action, pumping thermal (blue) and kinetic (red) energy into the
  ICM, increasing the gravitational potential energy (green) of ICM
  gas lifted out of the cluster centre and leading to an overall gain
  in the total energy (black). {\it Bottom panel:} run without
  radiative cooling. The total energy (black) within the central $125$
  kpc of the cluster increases in line with the energy injected by the
  jet (grey, $E_{\rm Inj}$), dominated by a gain in thermal (blue)
  energy, while the smaller gains in kinetic and gravitational
  potential energy reach similar levels to each other by $\sim 45$
  Myr.}
\label{fig:jet_energy_budget}
\end{figure}

The impact of the different (de-)refinement schemes on the physical
properties of the jet is further illustrated in Fig.
\ref{fig:jet_overview_deref}.  The bottom half of the top row shows
density and temperature slices through the $y=0$ plane for a kinetic
jet at $t\simeq 45$ Myr, for the SLR, JFD and GD schemes, from left to
right, respectively. The bottom row provides a quantitative overview
for the evolution of jet properties, showing (from left to right) jet
length, mass components and energy content for the SLR (blue), JFD
(red) and GD (green) schemes. The jet properties are calculated as
outlined at the end of $Section$ \ref{sec:sim_jet_props}.  For comparison,
the total injected jet energy is
\begin{equation}
  E_{\rm Inj}(t) = \int_{t_0}^{t}\dot{E}_{\rm J}(t'){\rm d}t',
  \label{eq:e_inj}
\end{equation}
where $\dot{E}_{\rm J}$ is taken from equation
(\ref{eq:jet_energy_rate}) and is shown by the solid black line in the
lower right-hand panel.

As already discussed, the morphology of the produced jets and their
ability to propagate depends greatly on the de-refinement scheme. With
strong de-refinement, the momentum (and energy) injected into the jet
is diluted in more massive cells and hence the jet growth is
stunted. However, as the occurrence of the de-refinement is reduced,
the jets are able to travel further in the
$z$-direction. Qualitatively, one can also see from the panels in the
top row of Fig. \ref{fig:jet_overview_deref} that stronger
de-refinement can smooth out instabilities along the jet--ICM
interface, potentially leading to poorly modelled mixing.

It is instructive to split the gas into different mass components, and
thus, we define the total mass of jet material within cells as $$M_{\rm
  J}=\sum_{f_{\rm J}>0}f_{J}m_{\rm cell},$$ the total mass of jet
material within the jet lobes as 
$$M_{\rm J, (lobe)}=\sum_{f_{\rm J}>0.01}f_{J}m_{\rm cell},$$ and the
total lobe gas mass as $$M_{\rm lobe}=\sum_{f_{\rm J}>0.01}m_{\rm
  cell}.$$ The evolution of these masses is shown in the lower middle
panel of Fig. \ref{fig:jet_overview_deref} by the dotted, dashed
and solid lines, respectively. We see that while all three runs
contain very similar masses of jet material in total, the GD scheme
has the highest mass in jet lobe material and retains the most jet
material within the jet lobes. Further, considering the energy content
(bottom right-hand panel), we find that while facilitating the
longest and most massive jet, the GD scheme also facilitates the jet
that retains the most injected energy ($\sim 30$ per cent of $E_{\rm Inj}$
in material with $f_{\rm J}>0.01$). On the other hand, while the jet
in the SLR scheme is shorter than that of the JFD scheme, it retains
more mass and energy.

Given the similar total injected jet mass between the schemes, both the jet
lobe mass and the jet mass within the lobes provide an
indication of the level of mixing (physical and numerical) of jet
material with the ambient gas. We can see from the bottom
middle panel of Fig. \ref{fig:jet_overview_deref} that the JFD
scheme retains the least jet material and entrained material within
the jet lobes, followed by the SLR scheme and then the GD
scheme. Which suggests most efficient mixing between the lobe material
and ambient gas in the JFD scheme and least efficient in the GD
scheme. Therefore, while we may have naively expected the strongest
de-refinement scheme (SLR) to result in the highest mixing, we see
that the JFD scheme is most efficient at mixing jet lobe material with
the ICM. We suggest that this is because of the larger surface area
for mixing produced with the JFD scheme compared to the pure SLR scheme.
		
A further consideration is the level to which one should refine cells
in the first place.  In \citet{CurtisSijacki15}, the purpose of the
model is to resolve the Bondi radius in order to more accurately model
accretion on to the BH and so they set the minimum cell radius to be
equal to the Bondi radius, i.e. $r_{\rm cell}^{\rm min}=r_{\rm
  Bondi}$. We tested the suitability of this choice and find that this
provides sufficient initial resolution for jet injection. Perhaps more
critically, we have tested parameters used for the additional JR scheme (see Section \ref{sec:refinement}). In addition to the
fiducial values of $\alpha = 0.01$ and $\gamma = 3$, we tried $\alpha
= 0.1$ with $\gamma = 3$ and $\alpha = 0.01$ with $\gamma = 2$ and
$4$. These are shown in Appendix \ref{App:JetRef}, with generally good
convergence of jet properties.

\subsubsection{Energy and/or momentum injection}
\label{sec:energy_mom_comp}

\begin{figure*}
\psfig{file=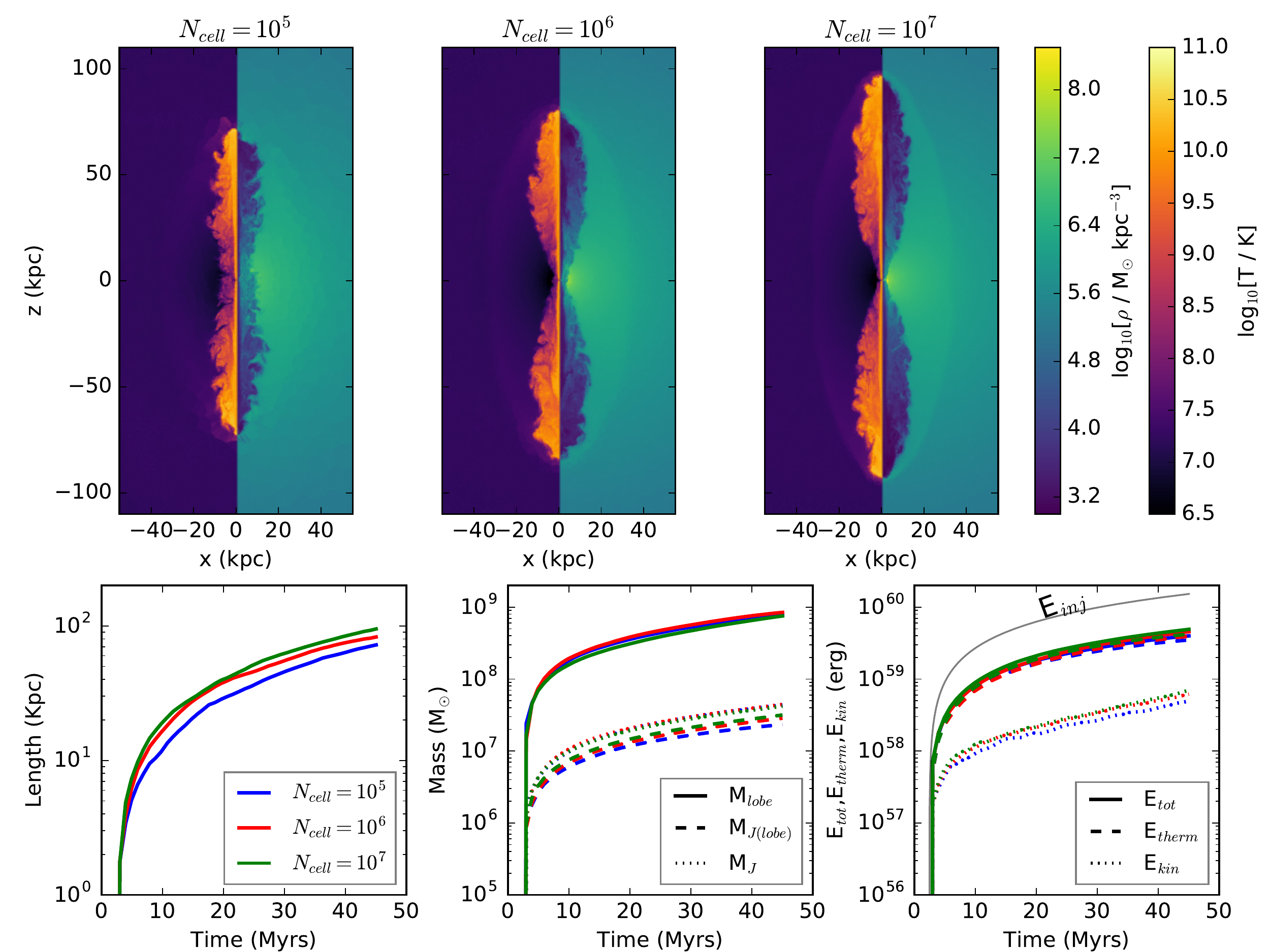,
  width=1.\textwidth,angle=0}
\caption{Overview: dependence of jet evolution (for the kinetic jet model) on global
  resolution of the simulation. The jet properties remain remarkably
  consistent over three orders of magnitude in mass resolution. Top row: density and temperature slices through the $y=0$ plane
  at $t\simeq 45$ Myr. Bottom row: evolution of the jet length
  (left-hand panel), different components of jet mass (middle panel)
  and different components of jet energy content (right-hand panel)
  for the kinetic runs with $N_{\rm cell}=10^{5}$ (left-hand panel and
  blue curves), $10^{6}$ (middle panel and red curves) and $10^{7}$
  (right-hand panel and green curves). For comparison, we show the
  total injected jet energy (equation \ref{eq:e_inj}) by the solid
  black line in the lower right-hand panel.}
\label{fig:jet_overview_res}
\end{figure*}

As discussed in Section \ref{sec:sim_jet_props}, we have implemented
different jet models with regard to momentum and energy injection. In
this section we outline how these models impact the evolution of jet
properties. The top row of Fig. \ref{fig:jet_overview_norm} shows
density and temperature slices at $t\simeq 45$ Myr for kinetic,
thermal and momentum jet runs, from left to right, respectively. The
bottom row shows the evolution of jet length, mass and energy in
different components for these three runs. Differences in morphology
due to the different injection schemes are clearly visible in the
panels in the top row. In terms of morphology, the kinetic and
momentum runs are similar, with a double lobe structure. Although the
size of the jet in the momentum run is much smaller because only $\sim
23$ per cent of the jet energy has actually been injected by this time (in
other words, $77$ per cent of the jet energy is explicitly lost due to mass
loading and momentum cancellation). The jet structure in the thermal
run includes an inflated central region, due to the expansion of the
gas when thermal energy is injected, and is shorter than the other
jets. Additionally, although the thermal jet has a greater energy
content than the momentum jet, it is much shorter because the
additional thermal energy results in a broader jet and hence the jet
feels an increased ram pressure force acting against it, as discussed in Section
\ref{sec:analytical}.

If we consider the evolution of various mass components of these jets,
we see that unlike in Fig. \ref{fig:jet_overview_deref}, the
different energy injection techniques have different total jet {material} masses,
$M_{\rm J}$. This can be explained by remembering that $f_{\rm J}$ is
set to $1$ for cells in the jet cylinder, such that the mass of jet
material in those cells is set to $m_{\rm J}=m_{\rm cell}$. Therefore
the more often a cell that already contains jet material is injected
with further jet material within the jet cylinder, the lower the total
jet mass will be due to double counting. Therefore, one may expect
that jets that are less able to quickly push cells away from the BH's
location would {more often} inject already jet-rich cells with
new jet material. In line with this expectation, the total jet mass
increases with the jet length, such that the kinetic jet run contains
the highest mass of jet material and the thermal jet the
least. Correspondingly the mass of jet material within the jet lobes
and the total mass of the jet lobes follow similar trends.

Considering the bottom right-hand panel, the kinetic and thermal jet
lobes contain similar amounts of energy, while the energy in the
momentum run is reduced by a factor of $\sim 5$, as expected from energy
conservation. Given both the momentum and thermal jets intrinsically
conserve momentum, they have similar kinetic energy content, while the
kinetic jet has a higher kinetic energy due to the explicit momentum
boost it received ($\sim 2\times\int\dot{p}_{\rm J}{\rm d}t$). Despite this,
the kinetic energy contributes $\simlt 13.5$ per cent of the total lobe
energy in all cases. Therefore, even if we assume that all of this is
turbulent kinetic energy (which is unlikely due to the large bulk
velocity of the jet), that places an upper limit of the turbulent
energy within the lobes at the level of a few percent when compared to
the total injected energy.

Specifically considering the energy budget of the kinetic jet, the
total lobe energy for material with $f_{\rm J} > 0.01$ is $\sim 30$ per cent
of the total injected jet energy. This rises to $\sim 40$ per cent if we
include gas with $f_{\rm J} > 0.001$ and is consistent with other simulations of jet lobe inflation that find,
  depending on jet parameters, $\sim 40$--$60$ per cent of the energy is
  retained in the jet lobes \citep[e.g.][]{HardcastleKrause13,
    EnglishEtAl16, WeinbergerEtAl17}. In Fig. \ref{fig:jet_energy_budget}, we plot how the total energy budget for
gas, minus the energy content when the jet is first activated, evolves during the injection of a
kinetic jet. We only consider gas within a sphere whose mass is equal to the total mass of gas ($\sim 3.1\times 10^{12}$ M$_{\odot}$) within the central $125$ kpc of the cluster at $t=0$. The top panel shows the run with radiative cooling, while
for comparison, the bottom panel shows a run without radiative
cooling. Solid curves show gains in energy, while dashed curves show
losses. In the non-radiative run, we see that the total energy (black
curve) within the central region increases by the amount injected by
the jet and is dominated by the thermal component (blue curve, $\simeq
0.64\times E_{\rm Inj}$), followed by the kinetic energy (red curve, $\simeq 0.2\times E_{\rm Inj}$) and
gravitational potential energy components (green, $\simeq 0.16\times E_{\rm Inj}$) by
$t\simeq 45$ Myr. Evidently, a sizeable portion of the injected
energy goes into driving the expanding cocoon and lifting ICM material
out of the gravitational potential well of the galaxy
cluster. However, {\it without} radiative cooling thermal energy dominates the budget at this point, with this energy split roughly equally between lobe and ICM gas. When radiative cooling is included (top panel), $\simeq 0.18\times E_{\rm Inj}$ goes into the kinetic energy component,
while the total change in gravitational potential energy is much
lower than in the non-radiative run, which peaks at $\simeq 0.07\times E_{\rm Inj}$ by $\simeq 34$ Myr but drops to $\simlt 0.04\times E_{\rm Inj}$ by $t\simeq 45$ Myr. The decrease in gravitational potential energy after $\simeq 34$ Myr indicates that after this point in time, gains in gravitational potential energy due to gas being pushed to larger radii are outweighed by losses in gravitational potential energy due to gas flowing back into the potential well of the cluster. Despite the significant amounts of energy being injected by the jet, there is still a net loss of thermal energy within the ICM, in the radiative run. However, globally, these losses are outweighed by the total energy injected by the jet, with
the system gaining in total energy at a rate of $\simeq 0.16\times\dot{E}_{\rm J}$. 

\begin{figure*}
\psfig{file=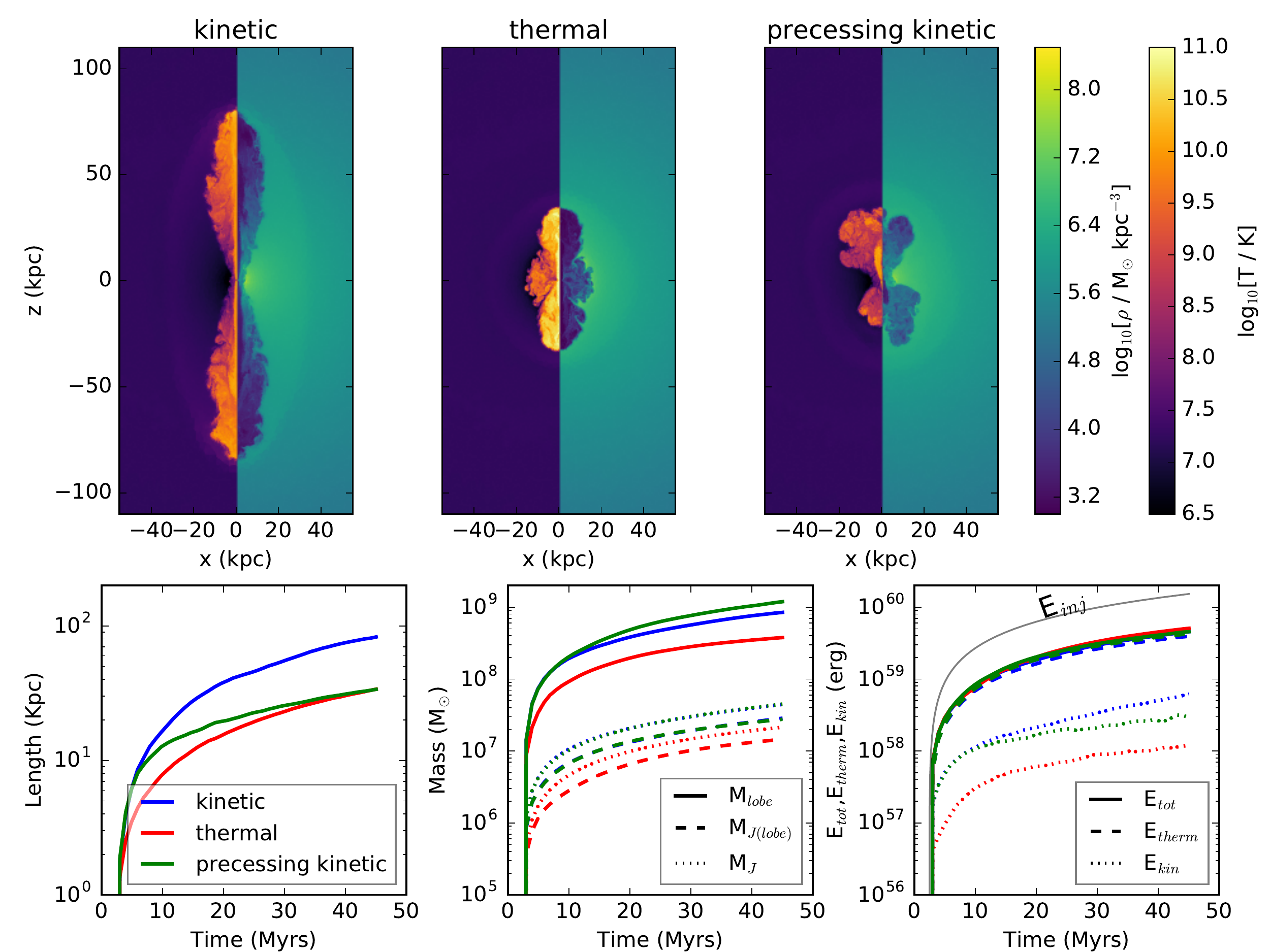,
  width=1.\textwidth,angle=0}
\caption{Overview: dependence of jet evolution when jet
  precession is included. The precessing jet shows a mass and energy
  contents similar to the kinetic jet, however a morphology closer to
  that of the thermal jet model. Top row: density and
  temperature slices through the $y=0$ plane at $t\simeq 45$ Myr. Bottom row: evolution of the jet length (left-hand panel), different components of jet mass (middle panel) and different
  components of jet energy content (right-hand panel) for the kinetic
  (left-hand panel and blue curves), thermal (middle panel and red
  curves) and precessing (right-hand panel and green curves) jets. For
  comparison, we show the total injected jet energy (equation
  \ref{eq:e_inj}) by the solid black line in the lower right-hand
  panel.}
\label{fig:jet_overview_prec}
\end{figure*}

\subsubsection{Resolution}

Finally, we consider the impact of changing the global resolution of
the simulation by performing additional {kinetic jet} runs with $N_{\rm
  cell}=10^{5}$ and $10^{7}$, giving corresponding target cell masses
of $m_{\rm cell}=1.8\times 10^{8}$ and $1.8\times 10^{6}$ M$_{\odot}$,
respectively. The initial conditions are set-up as described in Section
\ref{sec:sim_setup}. Fixing the mass, opposed to the
  number of neighbours, into which feedback energy is injected can
  provide better numerical convergence \citep[e.g.][]{BourneEtAl15}
  and so between resolutions $h_{\rm BH}$ is set such that $\sum
m_{\rm cell}(r) = 5.76\times 10^{8}$ M$_{\odot}$ for $r < h_{\rm
  BH}$ and $M_{\rm Jet}=10^4$ M$_{\odot}$ in all runs. 
  As in previous figures, we plot density and temperature slices
in the top row of Fig. \ref{fig:jet_overview_res}, with improving
resolution from left to right, respectively, and in the bottom row plot
the evolution of jet properties. Although the jet morphologies are
quite similar, improved resolution does result in slight increases in
jet length. It is also clear from the slices that larger instabilities{, in the form of physically larger Kelvin--Helmholtz eddies,} are observed in the lowest resolution run, which we would expect to result in increased mixing. This is born out if we compare the
fraction of injected jet material that remains within the jet lobes
($f_{\rm J} > 0.01$), which decreases from $\sim 76$ per cent at the highest
resolution to $\sim 53.5$ per cent in the lowest resolution, while there is
also a difference of up to $\sim 23$ per cent in the jet lobe energy. This
indicates that at lower resolution more of the jet material mixes
with the (shocked) ICM material. The increased mixing at low resolution
is of a numerical nature and therefore unphysical. However, despite
these differences in mixing, the evolution of jet properties agree
remarkably well and are converging with increasing
resolution. We also note that if we instead consider
  material with $f_{\rm J} > 0.001$, the differences are less stark,
  with the fraction of jet material within this gas only ranging
  between $\sim 86$ and $92$ per cent between the lowest and highest resolutions
  and with only a negligible difference in the total energy within such
  material.

\section{Results: Cavity inflation, gas flows and precessing jets}
\label{sec:prec_jets}

In this section, we consider in more detail the jet inflation process
and subsequent evolution of gas flows in the vicinity of the jet. On
top of this, based on the theoretical suggestion that jets need to
precess in order to isotropically distribute energy to the ICM
\citep[e.g.][]{VernaleoReynolds06, Falceta-GoncalvesEtAl10,
  LiBryan14,YangReynolds16a,YangReynolds16b} and based on
observational evidence that {some} jets do indeed seem to precess (or at the
very least move) \citep[e.g.][]{DunnEtAl06, Marti-VidalEtAl11,
  BabulEtAl13, AaltoEtAl16}, we also consider the impact of a
precessing jet and compare the properties to those of the kinetic and
thermal jets we have presented previously.

\subsection{The precessing jet}

The precessing jet runs are identical to the kinetic jet runs except
that the jet axis precesses about the $z$-axis. As in
\citet{YangReynolds16a} we use an angle of $15^{\circ}$ and a period
of $\sim 10$ Myr. The top row of Fig. \ref{fig:jet_overview_prec} is
similar to previous figures, showing density and temperature slices at
$t\simeq 45$ Myr for kinetic, thermal and precessing jets, from left
to right, respectively. The bottom row shows the evolution of jet
properties. Comparison between the thermal and kinetic jets has
already been made in Section \ref{sec:energy_mom_comp}, however,
considering the precessing jet (right-hand panel), it exhibits a
morphology closer to that of the thermal jet (despite being injected
as a kinetic jet), but with a lower temperature, more comparable to
the kinetic jet.

Initially, the precessing jet length evolution follows that of the
kinetic jet (see lower left-hand panel); however, before the jet axis
has completed one rotation, the length evolution
flattens and tends towards that of the thermal jet. Interestingly, the
precessing jet retains a similar mass in jet material within the jet
lobes as the kinetic jet, although has a somewhat larger lobe mass
overall, indicating that it has entrained more ICM material. Finally,
the precessing jet lobe material retains a broadly similar energy
content to the other methods. More specifically, we find that the
kinetic energy content of the precessing jet sits in between the
thermal and kinetic jets, indicating that the precession of the jet
leads to a larger fraction of the initial kinetic energy of the jet
becoming thermalized when compared to the kinetic jet.

\begin{figure*}
\psfig{file=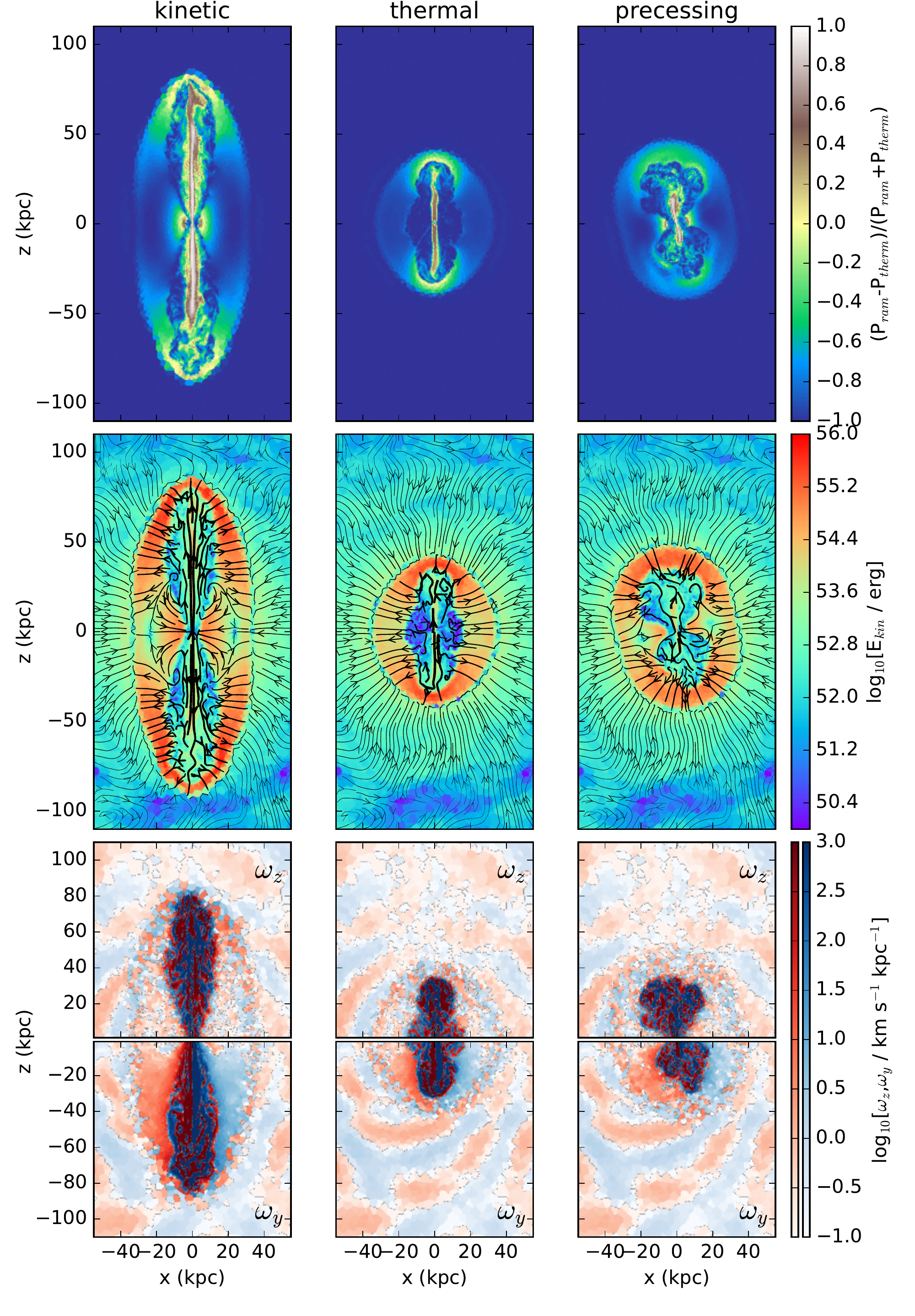,
  width=0.85\textwidth,angle=0}
\caption{Overview: comparison of gas flows and cocoon structure
  properties for the kinetic, thermal and precessing jet models. All
  panels show slices through the $y=0$ plane at $t\simeq 45$ Myr. Ram
  pressure dominated jets, which have large velocities and hence high kinetic energies, thermalize through shocks and inflate thermal pressure dominated, turbulent
  lobes. The lobes expand and drive a massive outflowing shell that defines
  the boundary of the cocoon. Top row: the ratio $f_{\rm
    P}=(P_{\rm ram}-P_{\rm therm})/(P_{\rm ram}+P_{\rm therm})$,
  indicating the relative contributions of ram and thermal pressure to
  the total pressure. Middle row: the gas {cell} kinetic energy is
  shown by the colour map, while the gas velocity field is shown by the
  overlaid streamlines, the thickness of which vary with $|{\textbf{\textit v}}|$. Bottom row: component of vorticity in the $z$- and $y$-direction, as labelled, with red and blue colours corresponding to oppositely
  directed vorticity vectors.}
\label{fig:jet_flows_prec}
\end{figure*}

\subsection{Jet inflation and gas flows}
\label{sec:jet_inflation_and_gas_flows}

Fig. \ref{fig:jet_flows_prec} shows slices of various jet properties
for the kinetic, thermal and precessing jet runs, from left to right,
respectively, at $t\simeq 45$ Myr. The top row illustrates the
dominant pressure component within the jet cocoon. As discussed in Section
\ref{sec:analytical}, the pressure within the cocoon plays an
important role in determining its expansion and the propagation of the
jet. The total gas pressure, $P_{\rm tot}$, is made up of a thermal
($P_{\rm therm}$) and a ram pressure ($P_{\rm ram}$) component. To
determine which dominates we define the parameter:
\begin{equation}
f_{\rm P}=\frac{P_{\rm ram}-P_{\rm therm}}{P_{\rm ram}+P_{\rm therm}},
\label{eq:pressure_ratio}
\end{equation}
such that $f_{\rm P}=1$ for ram pressure dominated gas and $f_{\rm
  P}=-1$ for thermal pressure dominated gas. This quantity is
displayed in the top row of Fig. \ref{fig:jet_flows_prec}. Clearly the jet itself is dominated by ram pressure with the
poles of the jet cocoon in the vicinity of the bow shock, also having a
large ram pressure contribution. On the other hand, the cocoon in
general and its expansion is dominated by thermal pressure. In
particular, as expected from early analytical models
\citep[e.g.][]{BlandfordRees1974, Scheuer1974, Begelman89}, the fast
jet is shock heated. This results in the lobes of shocked jet material
encapsulating the jet itself in which $f_{\rm P}\simeq -1$. This gas
expands due to the thermal pressure, driving the growth of the cocoon
and resulting in the shell of gas out to the edge of the cocoon for
which $-1 < f_{\rm P} < 0$. We further note that similar to
observations \citep[e.g.][]{FabianEtAl06, FormanEtAl07, CrostonEtAl11,
  SandersEtAl2016}, the expansion of the jet cocoon mainly drives only
weak shocks into the ICM, predominantly at the poles of
  the cocoon, where the expansion velocity exceeds the ICM sound speed,
  $c_{\rm s, ICM}$. While the cocoon expansion {\it perpendicular} to the
  jet direction also initially drives a shock wave into the ICM, after
  only a few Myr the typical radial velocity of cocoon material drops
  below $c_{\rm s, ICM}$ and the shock wave becomes a sound wave {\citep[see also][]{GuoEtAl17}},
  which propagates into the ICM with a velocity of $c_{\rm s, ICM}$. 

The middle row of Fig. \ref{fig:jet_flows_prec} shows {Voronoi cell} kinetic energy
slices overlaid with flow lines showing the velocity field in the
$y=0$ plane. The thickness of the flow lines scales logarithmically
with the gas velocity. The kinetic energy is greatest along the jet
axis and in the shell of cocoon gas, which, as shown by the
streamlines, is expanding into the ICM. {In the case of the kinetic jet,} we estimate that the kinetic
energy within the expanding shell accounts for $\sim 10$ per cent of the
total energy injected by the jet. A further point to note is that while the kinetic energy of the jet is dominated by high velocities, the kinetic energy of the expanding shell is dominated by its
mass. With regard to the velocity field, there is a clear,
elliptically shaped discontinuity at the location of the
  bow shocks and sound wave which outline the cocoon.  The
streamlines immediately beyond the cocoon boundary are inflowing,
because the gas is in the cooling flow region. The streamlines also
highlight interesting flow features around the base of the jet,
particularly in the case of the kinetic jet. The inflation of the jet
lobes and expansion of the cocoon displaces ambient ICM
gas that was initially in hydrostatic equilibrium. The displaced
material is pushed into a shell around the jet
lobes, some of which then falls towards the centre of the
cluster, converting gravitational potential energy into kinetic energy
in the process. This ultimately results in the build-up of a reservoir
of gas around the BH, providing further fuel for it, but also somewhat
suffocating the jets progress. Inspection of the velocity field shows
gas flows within the cocoon whereby material is dragged up by the jet,
circulates and returns back to the base of the jet with the inflowing
material. We note that an additional class of gas flows, dubbed
backflows in the literature \citep[e.g.][]{ADS2010, CieloEtAl2014}
are also observed in our simulations; however{, they are susceptible to disruption by turbulent motions within the jet cocoon, especially at later times.}

\begin{figure*}
\psfig{file=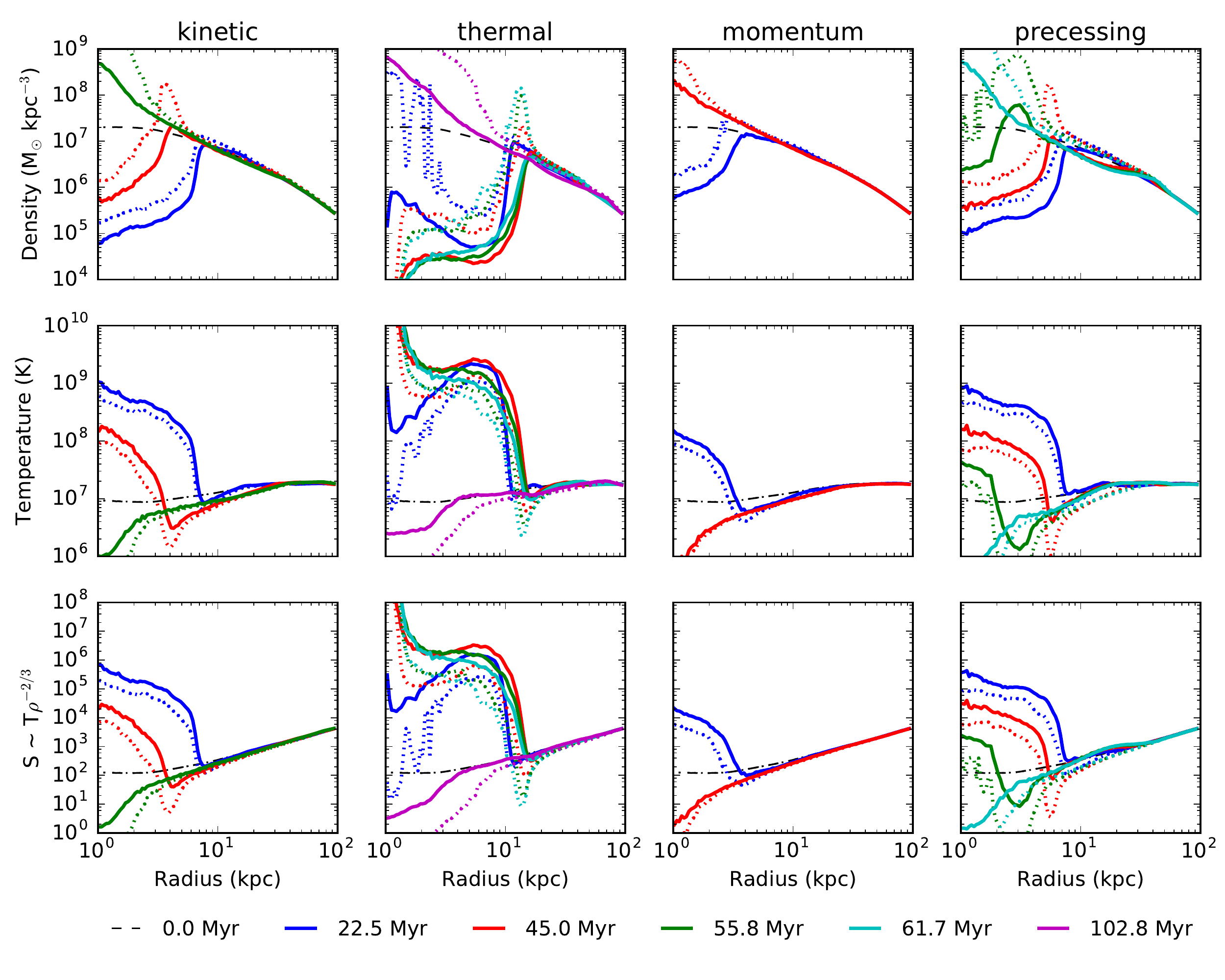,width=1.\textwidth,angle=0}
\caption{Overview: evolution of radial density ({top row}),
  temperature ({middle row}) and entropy ({bottom row})
  profiles for (starting from the left-hand column) a kinetic,
  thermal, momentum and precessing kinetic jet, respectively. {The density profile is calculated using either volume weighting (solid) or emission weighting (dotted), while the temperature profile is calculated using either mass weighting (solid) or emission weighting (dotted). The entropy profile is produced by combining the density and temperature profiles.} The inflation of
  the jet cocoon results in initially reduced central densities and
  increased temperature and entropy profiles. However, given that
  these jets have a fixed $\dot{m}$, the power input does not respond
  to the ICM properties. Thus, eventually material can flow back into
  the centre of the cluster, resulting in increased central densities
  and reduced central temperature and entropy profiles at later
  times. These panels should not be used as a direct comparison to
  observations because of the fixed jet power, but as a diagnostic of
  the different jet injection methods.}
\label{fig:norm_halo_props}
\end{figure*}

Finally, the injection of the jet into the ambient medium is expected
to drive some level of turbulence, while the propagation of the jet
may also lead to the formation of fluid instabilities. The vorticity
of the gas provides a good indicator of the production of such
turbulence. As such, the bottom row of Fig. \ref{fig:jet_flows_prec}
shows the $y$- and $z$-components of the vorticity generated after
$45$ Myr, in the bottom and top half of the jet, respectively. We
present these two components as they represent potentially different
modes of interaction of the jet with the ICM. As discussed in
\cite{ReynoldsEtAl2015}, in a stratified medium, the $x$- and
$y$-components of vorticity are primarily produced by $g$-modes,
whereas the $z$-component of the vorticity is a product of jet-driven
turbulence. It is clear from all three panels that the jets are able
to generate vorticity (both $y$- and $z$-components), with increases
of up to a factor of $\sim 1000$ compared to the initial vorticity of the
gas. The most significant generation of vorticity is confined to the
jet and shocked jet material, indicating that the jet can readily
drive turbulence and the production of $g$-modes here \citep[see also
][]{WeinbergerEtAl17}. To a lesser extent the jet is also able to
drive vorticity in the (shocked) ICM gas, as most clearly illustrated
for the kinetic jet in the lower left-hand panel. In particular, for
$\omega_{\rm y}$, there are large regions of coherent vorticity in the
(shocked) ICM material, suggesting the production of g-modes
here. However, beyond the cocoon there is no noticeable increase in
vorticity, suggesting that the g-modes are trapped within
  the cocoon boundary and that the jet is unable to drive large-scale turbulence{. This is} consistent with observations of the Perseus cluster by Hitomi \citep{Hitomi2016} and as found in other simulations
\citep[e.g.][]{ReynoldsEtAl2015, YangReynolds16a, YangReynolds16b,
  WeinbergerEtAl17}. We refer to Section
\ref{sec:substruct_vs_jet_turbulence} for a detailed comparison with
Hitomi observations.

{Additionally, while non-zero vorticity is the product of the incompressible component of the velocity field, a non-zero velocity divergence is the result of a compressible component, associated with shocks and sound waves \citep[e.g.][]{ReynoldsEtAl2015}. We can consider the relative importance of these components in different locations by considering the compressive ratio \citep[e.g.][]{IapichinoEtAl11}:
\begin{equation}
r_{\rm cs} = \frac{\left<|\nabla\cdot{\bf v}|^{2}\right>}{\left<|\nabla\cdot{\bf v}|^{2}\right> + \left<|\nabla\times{\bf v}|^{2}\right>},
\end{equation}
where $<...>$  represent mass-weighted averages performed over all Voronoi cells within the region of interest. By definition, $r_{\rm cs}=1$ in regions with purely compressible flow, i.e. $\nabla\times{\bf v}=0$. On the other hand, $r_{\rm cs}=0$ in regions with purely incompressible flow, i.e. $\nabla\cdot{\bf v}=0$. Our analysis finds that for jet lobe material ($f_{\rm J} > 0.01$), $r_{\rm cs}\simeq 0.03$ and hence is dominated by the incompressible velocity component associated with turbulence. If instead we consider outflowing cocoon material ($v_{\rm rad}> 10$ km s$^{-1}$ and $f_{\rm J} < 0.01$), we find that $r_{\rm cs}\simeq 0.85$ and so the flow here is dominated by the compressible component of the velocity field associated with sound waves and shocks. Recent works have considered the role of AGN feedback in triggering star formation \citep[e.g.][]{GaiblerEtAl12,NayakshinZubovas12, Silk13,ZubovasBourne17}, with the nature of the velocity field playing a potentially critical role in determining whether star formation is enhanced or inhibited \citep{FederrathKlessen12, FederrathEtAl17}}

\subsection{How different jet models impact cluster properties}
\label{sec:evo_cluster_props}

Here, we outline the impact of jets on the gas within the cluster for
the four injection techniques discussed in the previous
sections. Fig. \ref{fig:norm_halo_props} shows radial {profiles for different halo properties: the top row shows density, calculated using volume (solid) or emission (dotted) weighting; the middle row shows temperature, calculated using mass (solid) or emission (dotted) weighting; and the bottom row shows entropy calculated as $T\rho^{-2/3}$. Profiles are shown at} various times for kinetic, thermal, momentum and precessing jets, from left to right, respectively. Also, shown is a black dashed curve indicating the
initial, spherically averaged radial distributions. We note that given we are injecting jets with a fixed $\dot{m}_{\rm J}$, there is no self-regulation of the feedback. This was done intentionally to
provide a clean comparison between the different jet energy injection
techniques and as such Fig. \ref{fig:norm_halo_props} should be seen
as a diagnostic of these techniques and not be used as a direct
comparison to observed cluster profiles. 

The introduction of the jet initially leads to the central density
dropping sharply and the temperature and entropy increasing,
particularly along the axis of the jet. However, eventually material
that is displaced by the inflation of the jet cocoon flows towards the
base of the jet leading to increased densities and lower temperatures
and entropies at small radii. The physical size of the regions
affected by the jet and when exactly material flows inwards depends
upon the jet method used and how effectively it initially pushes
material away from the central regions. We note that while such sharp
changes in cluster radial profiles are not observed in real galaxy
clusters, as mentioned previously, we deliberately inject a fixed,
high power ($\sim 10^{45}$ erg s$^{-1}$) jet, in order to assess its
ability to drive turbulence within the ICM. As such, we do not expect
to produce profiles that match observed cluster profiles.

Comparing the different jet injection methods, we {find} that unsurprisingly,
the least effective method is the momentum-driven jet,
which impacts the smallest region with respect to the other jet models
and exhibits the smallest rise in central gas
temperatures. Furthermore, by $45$ Myr a central density cusp forms
and the central temperature and entropy drops below the original
values. On the other hand, both the kinetic and precessing jets can
influence a larger central region and while exhibiting similar
temperature profiles to each other, the precessing jet is slightly
more effective at preventing gas inflow perpendicular to the jet
direction. The thermal jet run can affect the largest area of gas and
heats the gas to significantly higher temperatures than any of the
other jet models. Further, the thermal jet is the most effective at
preventing gas returning to and accumulating within the cluster
centre.

\begin{figure*}
\psfig{file=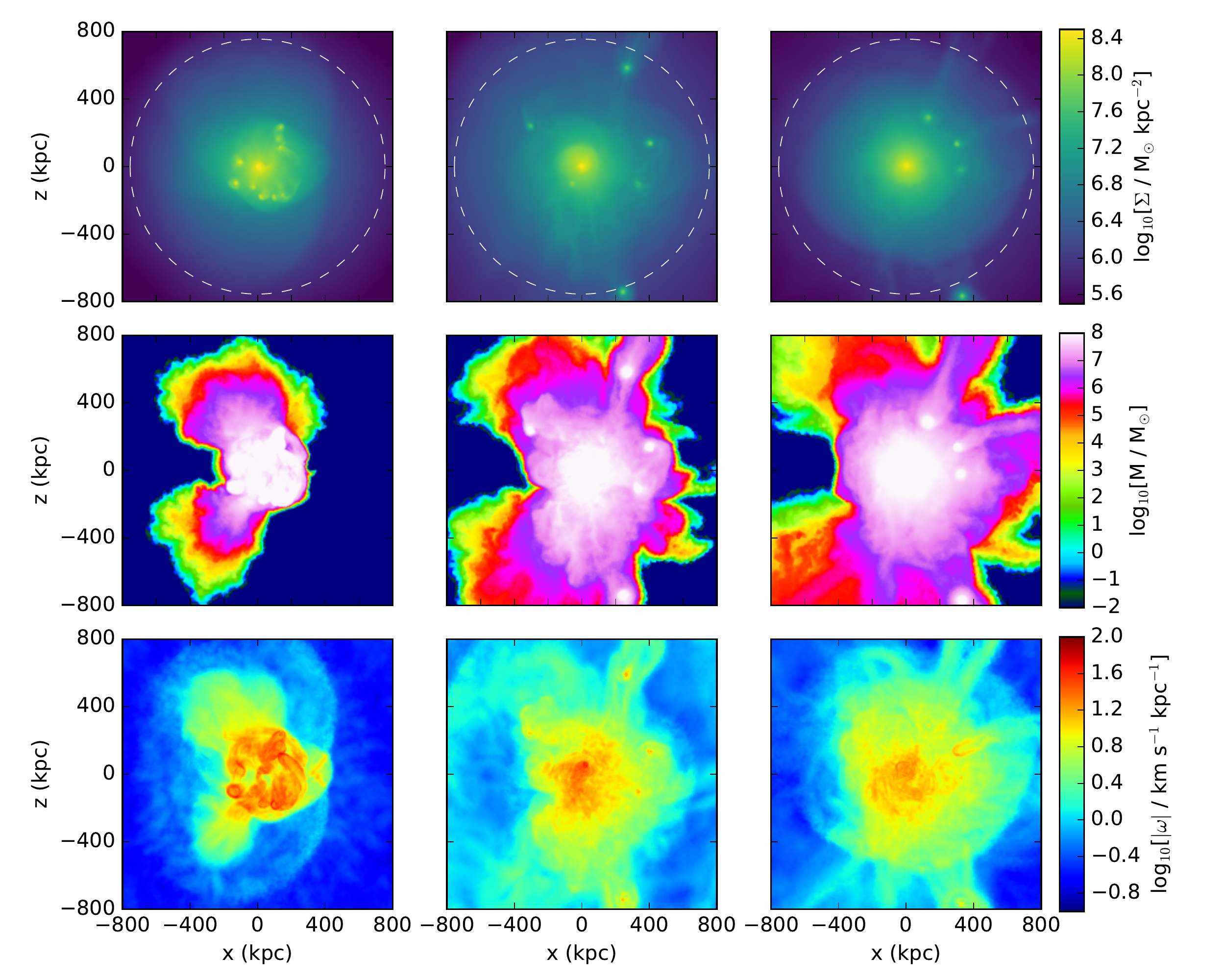,width=1.\textwidth,angle=0}
\caption{Overview: evolution of projected ICM properties stirred
  by the motions of substructures. The stirring results in density
  inhomogeneities, mixing of substructure and ICM material and the
  driving of a turbulent velocity field as illustrated by vorticity
  generation. The columns from left to right show projections after
  $\sim 1.25$, $2.25$ and $3.25$ Gyr of stirring by
  substructures. Top row: evolution of the gas column density,
  with $r_{200}$ indicated by the dashed white circle. {Middle
    row:} evolution of the projected subhalo gas mass, calculated as
  the $\int(f_{\rm sub}\rho_{\rm cell}){\rm d}z{\rm d}A$ along the
  line of sight. Bottom row: Evolution of the projected
  vorticity, calculated as a mass-weighted projection of $|{\bf
    \omega}|$.}
\label{fig:substruct_rho_proj}
\end{figure*}

The large difference between the thermal jet run and the other runs
can be attributed to the direction in which the feedback is acting. In
the kinetic and momentum runs, and broadly speaking in the precessing
run, the feedback is primarily directed along the jet axis, with only
shock {and compressional} heating providing more isotropic feedback. On the other hand,
for the thermal jet, the internal energy, which is explicitly injected,
leads to expansion of the heated gas more isotropically, thus more
readily clearing the central regions. 

\section{Results: substructure-driven turbulence and jets}
\label{sec:sub_structure}

\begin{figure}
\psfig{file=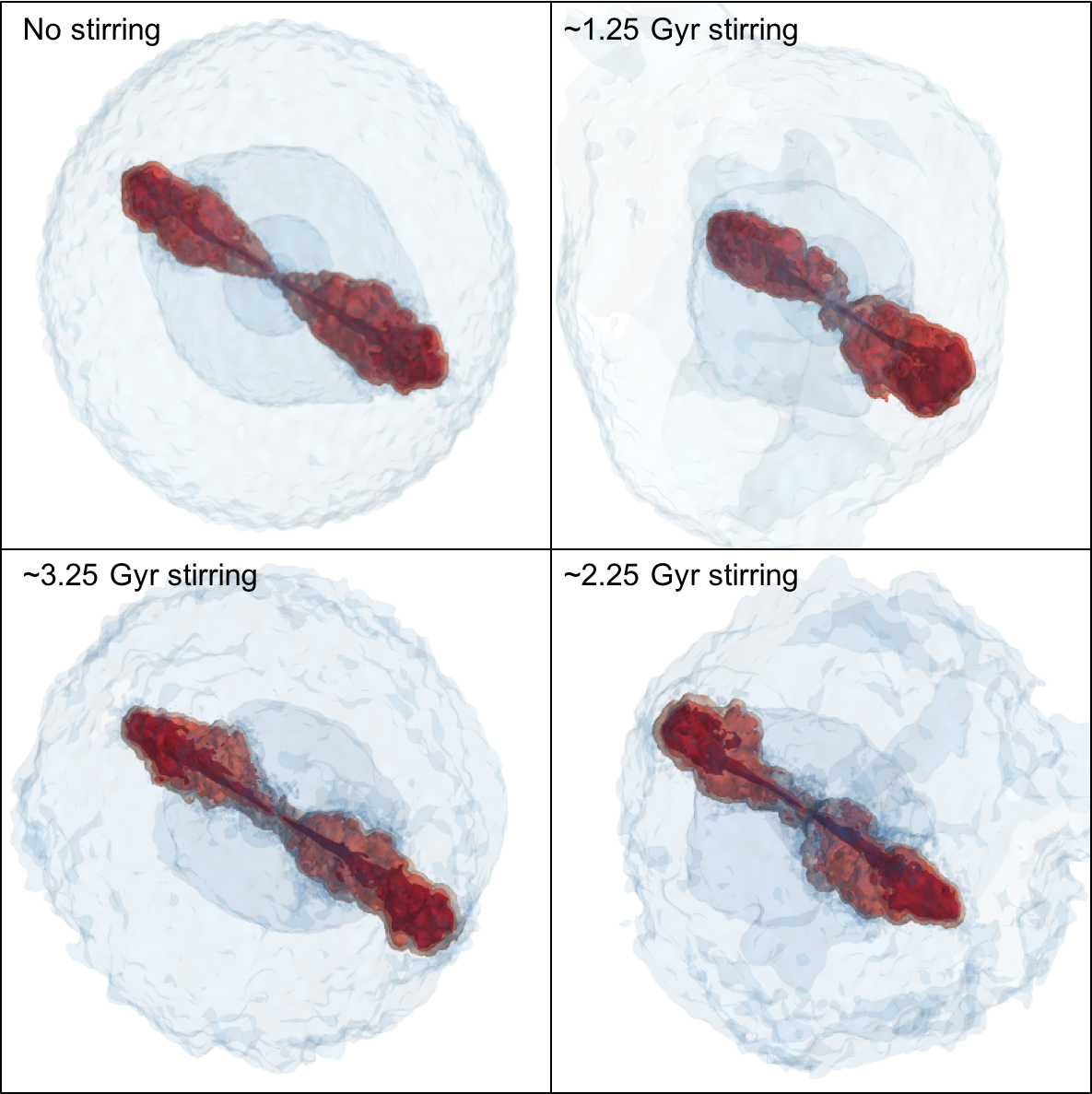,
  width=0.48\textwidth,angle=0}
\caption{Overview: illustration of jet morphologies when the jet
  is injected into a medium that has been stirred by the motion of
  substructures compared to a run in which there is no stirring (top
  left). The figure shows that the gas motions in the ICM can distort
  the jet lobe morphology to become rather asymmetrical. {Clockwise from top left:} $3$D ambient ICM density (blue) and hot
  gas temperature (red) contours for a jet with no stirring and jets
  injected into a medium that has been stirred by substructure motions
  for $\sim 1.25$, $2.25$ and $3.25$ Gyr, respectively. The total jet
  length (end to end) in the fiducial kinetic run is $\sim 180$ kpc.}
\label{fig:jet_contours_substruct}
\end{figure}

\subsection{The `stirred' ICM}

Up to this point, we have presented jet feedback in a homogeneous and
smooth ICM. However, realistic galaxy clusters are expected to contain
substructures which stir {the ICM} and possibly drive turbulence
\citep{DolagEtAl2005, VazzaEtAl12, GuEtAl2013,
  ZuHoneEtAl2013, VazzaEtAl17}. As such, it is interesting to investigate how {these}
motions impact the evolution of the jet and its cocoon. In this
section, we perform simulations with a fixed $\dot{m}_{\rm J}$ kinetic
jet, as described in Section \ref{sec:energy_mom_comp}, however, with
modified initial conditions. Specifically we adopt a set-up similar to
that presented in \citet{SijackiEtAl12} that includes ten $2\times
10^{12}$ M$_{\odot}$ subhaloes added to the relaxed initial conditions
described in Section \ref{sec:sim_setup}.  Each subhalo consists of a live
\citet{Hernquist90} dark matter potential consisting of $1000$ dark
matter particles and a gaseous component with gas fraction, $f_{\rm
  g}=0.17$. The subhaloes are randomly positioned between {$r = 625$ and
$775$ kpc} from the centre of the main halo, with their orbital
velocities set to between $200$ and $500$ km s$^{-1}$. This set-up is
run non-radiatively and without jet feedback to allow the motions of
the subhaloes to stir the ICM, inducing turbulence and bulk motions
within the main halo gas that have the potential to
impact jet evolution. Fig. \ref{fig:substruct_rho_proj} shows the
column density (top), subhalo tracer mass (middle panel) and
projected vorticity (bottom), respectively, at $t\simeq 1.25$, $2.25$ and $3.25$
Gyr (left to right, respectively). Also shown in the top panels is
$r_{200}=754$ kpc, indicated by the dashed white line.

As the subhaloes move through the ICM, their locations are shown at
different times in the top three panels of Fig.
\ref{fig:substruct_rho_proj}. The motion of the subhaloes stirs the
ICM, leading to the formation of inhomogeneous density structures and
streams as gas is stripped from the subhaloes. The three snapshots
provide quite different levels of perturbation and are each used as
initial conditions for jet simulations. The level of stripping and
subsequent mixing of the gas initially associated with the subhaloes
can be seen in the middle row, which shows the line-of-sight mass of the
material initially within the subhaloes.  We use a tracer field,
$f_{\rm sub}=m_{\rm sub}/m_{\rm cell}$, similar to that used for jet
material in which $f_{\rm sub}=1$ for the cells that initially make up
the subhaloes. The tracks of the stripped subhalo gas are shown by the
trails in the middle row, which shows the extent to which the material
permeates the central region of the cluster.

The impact of the motions of the subhaloes with regard to the velocity
field and the driving of turbulence can be seen in {the} bottom panels,
which show the projected absolute value of the gas vorticity
$|{\bf\omega}|$. Initially, the gas vorticity is negligible; however,
as the subhaloes move through the cluster, quite significant vorticity
is generated. {Out of the three new initial conditions,} the vorticity is strongest around $t\sim 1.25$ Gyr when
it is concentrated in the centre of the cluster during the subhaloes
first passage through this region (see bottom left-hand panel). As the
subhaloes then leave pericentre and travel at larger radii, vorticity
is produced on larger scales, but the magnitude subsides
a little with time (see bottom right two panels), but is still
strongest in the central regions of the cluster. The turbulence
produced by the motions of the subhaloes provides extra pressure
support to the cluster gas, leading to a reduced central density (for
further details see Section \ref{sec:evo_halo_sub}) and the motions
produced influence the evolution of jets, as we discuss now.

\subsection{Jets in a `stirred' ICM}

Now that we have outlined the evolution of the basic properties of the
ICM during the non-radiative substructure runs, we consider the
different jet runs. First, Fig. \ref{fig:jet_contours_substruct}
shows $3$D contours of ambient ICM density (blue) and hot jet gas
(red) for runs with and without stirring by substructure motions. From
the top-left and moving clockwise, the panels show kinetic jets at
$t\simeq 45$ Myr in a system where the ICM has not been stirred, and
has been stirred for $\sim 1.25$, $2.25$ and $3.25$ Gyr,
respectively. 

The first thing to notice is that the jets injected into a stirred ICM
are somewhat disturbed and less symmetrical than the fiducial kinetic
jets shown in the top-left panel, with the gas motions induced in the
ICM by the subhaloes having a noticeable impact on the jet
morphology. However, we find that the global jet lobe properties, such
as length, mass and energy content remain remarkably similar between
all of the runs. In this sense, it seems that there is no clear trend
as to how concurrent substructure motions impact mixing of the lobe
material, while the jet is active. Nonetheless, as we will show in Section
\ref{sec:flows_with_subs}, the motions can significantly disrupt the
(shocked) ICM cocoon material and potentially aid its mixing with the
ambient ICM. Although we should note that both in runs with and without substructures, the majority of jet
material remains within the jet lobes while the jet is active ($\sim 60$ per cent for $f_{\rm J} > 0.01$ and $\sim 90$ per cent for $f_{\rm J}>0.001$). Interestingly, however, in Section \ref{sec:sound_waves} we will show that after the jet has become inactive, substructure motions can have a significant impact on the disruption and mixing of jet material when compared to runs without substructures.

\begin{figure*}
\psfig{file=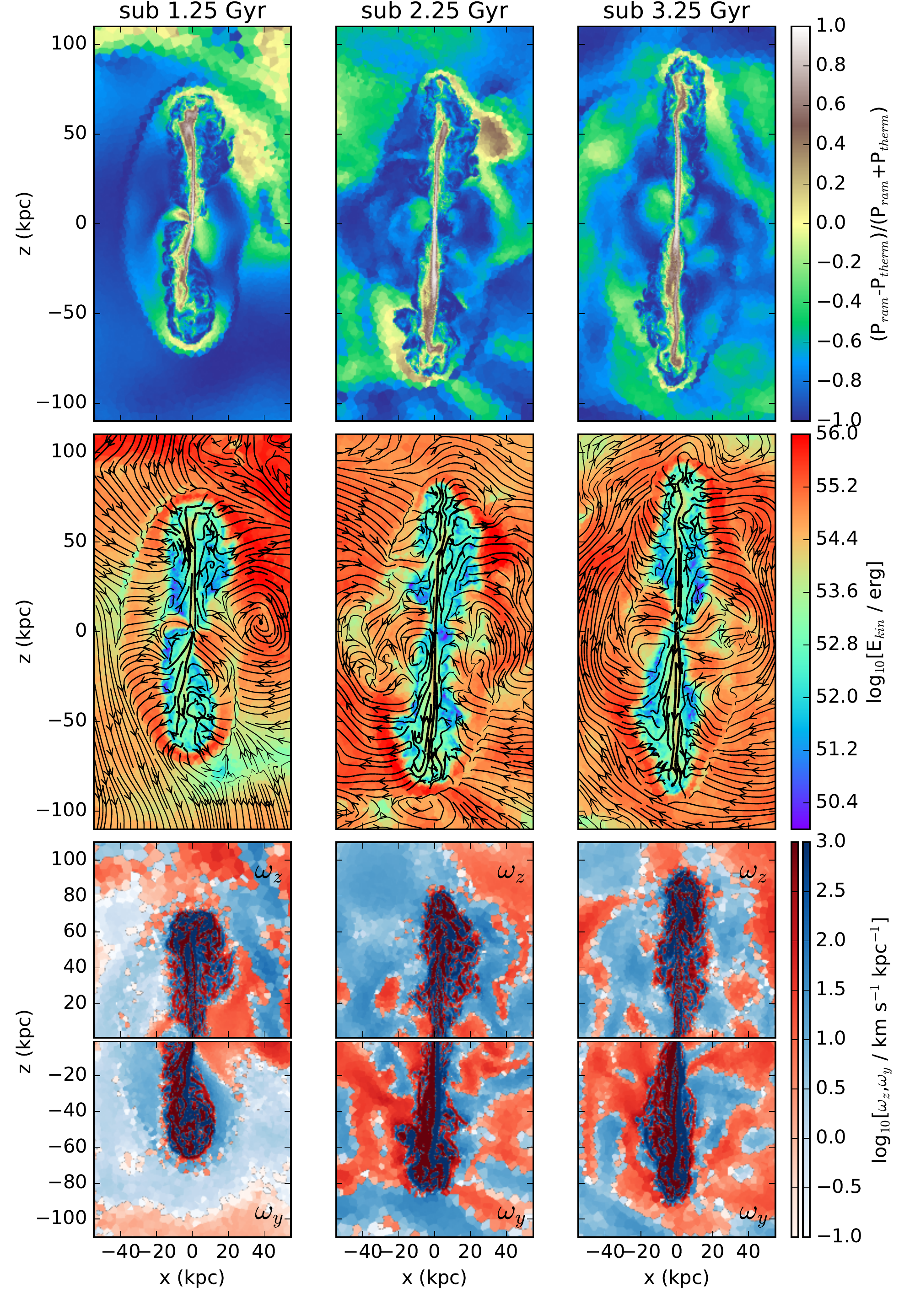,
  width=0.85\textwidth,angle=0}
\caption{Overview: comparison of gas flows and cocoon structure
  properties for a kinetic jet in a cluster that has been stirred for
  $\sim 1.25$, $2.25$ and $3.25$ Gyr by substructure motions. All
  panels show slices through the $y=0$ plane at $t\simeq 45$
  Myr. Unlike in Fig. \ref{fig:jet_flows_prec}, stirring by
  substructures results in a non-negligible level of ram pressure and
  vorticity within the ICM. These motions can disrupt and dominate the
  jet cocoon, resulting in a poorly defined outer boundary of the
  cocoon. Top row: the ratio $f_{\rm P}=(P_{\rm ram}-P_{\rm
    therm})/(P_{\rm ram}+P_{\rm therm})$, indicating the relative
  contributions of ram and thermal pressure to the total
  pressure. {Middle row:} The gas {cell} kinetic energy is shown by the
  colour map, while the gas velocity field is shown by the overlaid
  streamlines, the thickness of which vary with $|{\textbf{\textit v}}|$. Bottom row: component of vorticity in the $z$ and $y$ direction, as
  labelled, with red and blue colours corresponding to oppositely
  directed vorticity vectors.}
\label{fig:jet_flows_substruct}
\end{figure*}

\begin{figure*}
\psfig{file=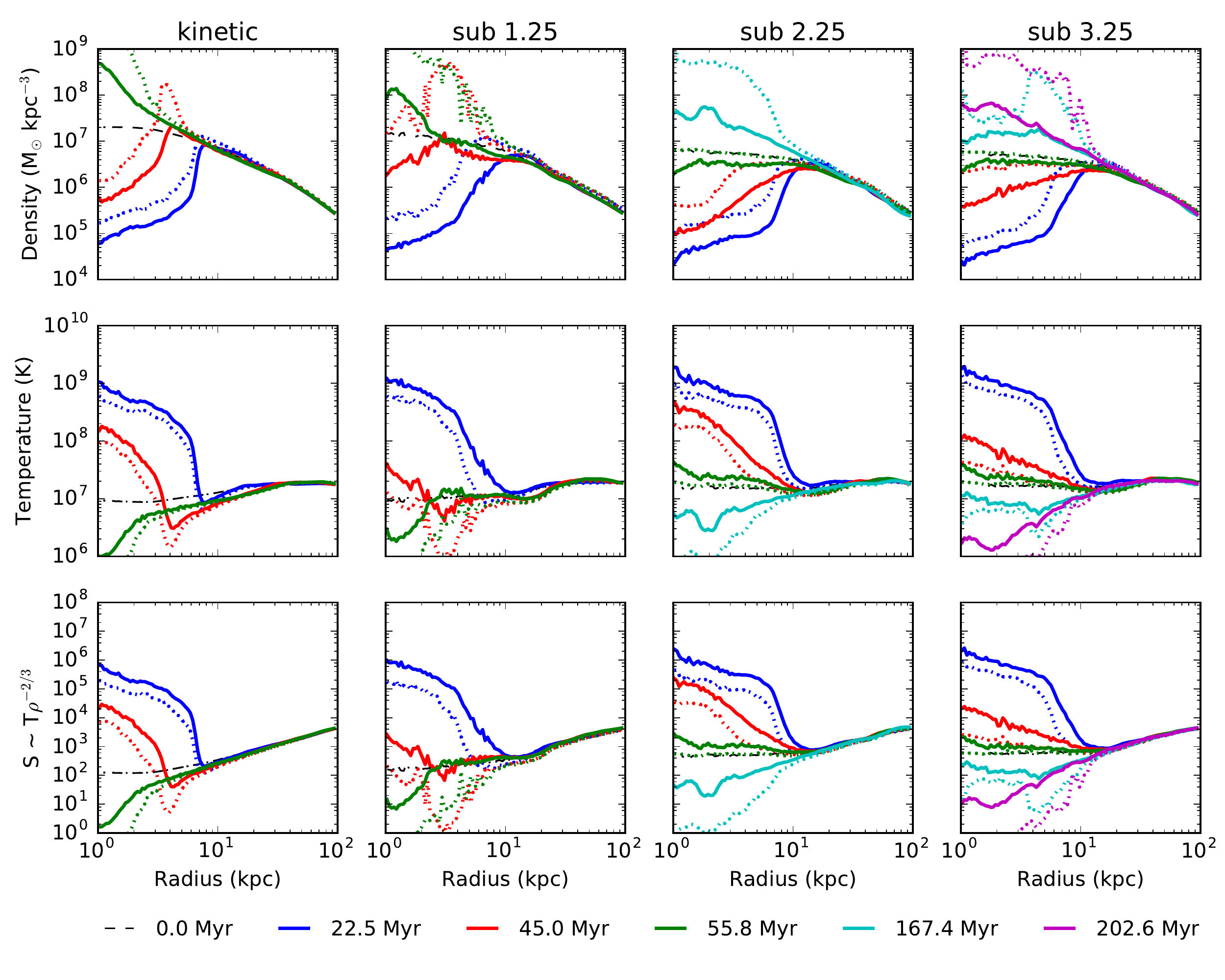,width=1.\textwidth,angle=0}
\caption{Overview: evolution of radial density ({top row}),
  temperature ({middle row}) and entropy ({bottom row})
  profiles for a kinetic jet in an ICM that is (starting from the
  left-hand column) not stirred, or stirred for $\sim 1.25$, $2.25$
  and $3.25$ Gyr, respectively. {The density profile is calculated using either volume weighting (solid) or emission weighting (dotted) while the temperature profile is calculated using either mass weighting (solid) or emission weighting (dotted). The entropy profile is produced by combining the density and temperature profiles.} The motions driven by the substructures provide additional pressure support to the ICM and prevent the infall of gas into the central region of the cluster. However, as in Fig. \ref{fig:norm_halo_props}, given that these jets have a fixed
  $\dot{m}$, eventually material does flow back into the centre of the
  cluster.}
\label{fig:substruct_halo_props}
\end{figure*}

\subsection{Jet inflation and gas flows}
\label{sec:flows_with_subs}

The relative importance of ram pressure and thermal pressure, as
encapsulated in the quantity $f_{\rm P}$ (see equation
\ref{eq:pressure_ratio}) is shown in the top row of Fig. \ref{fig:jet_flows_substruct}. The plot is in the $y=0$ plane at
$t\simeq 45$ Myr for the jets initiated after $1.25$, $2.25$ and
$3.25$ Gyr of substructure motions. Similar to the jets shown in the
top row of Fig. \ref{fig:jet_flows_prec}, the jet pressure is
dominated by the ram pressure component, while the lobe and cocoon
pressure is dominated by thermal pressure. In contrast to the fiducial
run, due to the subhalo-driven motions, the gas beyond the cocoon also
has a ram pressure component, which in some cases makes a contribution
comparable to that of the thermal component. Further, it is
interesting to consider the gas pressure components close to the base
of the jet. In the top left-hand panel of Fig.
\ref{fig:jet_flows_prec}, which shows the fiducial kinetic jet, the
pressure in the plane near the base of the jet is ram pressure
dominated. Combining this information with the field lines shown in
the middle left-hand panel of Fig. \ref{fig:jet_flows_prec}, it is
clear this is due to material flowing into the centre of the halo.  In
comparison, the central ram pressure component is reduced in the top
left-hand panel of Fig. \ref{fig:jet_flows_substruct}, while the
central pressure is dominated by the thermal component in the other
two top panels.  This shows that gas motions driven by subhaloes help
to prevent gas inflow within the innermost region of the cluster.

The impact of the gas motions produced by substructures can be seen in
the middle row of Fig. \ref{fig:jet_flows_substruct}, which like the
corresponding row of Fig. \ref{fig:jet_flows_prec} shows slices of
the {Voronoi cell} kinetic energy, with the velocity field indicated by overlaid
streamlines, the thickness of which scales logarithmically with the
gas velocity magnitude. A clear difference between the runs with and
without substructures is the additional kinetic energy beyond the jet
cocoon seen in the middle row of Fig. \ref{fig:jet_flows_substruct},
which is of a similar level to the jet cocoon material. In the
fiducial jet run, shown in Fig. \ref{fig:jet_flows_prec}, there is a
clear discontinuity in the velocity field indicating the ICM shock{/sound wave} at
the boundary of the jet cocoon. However, this can be disrupted by the
concurrent ICM velocity field in the substructure runs. In the left
most panel this influence only occurs in small regions of the jet
cocoon, where a vortex in the ambient gas clearly breaks into the
cocoon in the top right region. However, in the other two cases, the
velocity discontinuity is almost completely lost.

Finally, the bottom panels of Fig. \ref{fig:jet_flows_substruct} show
the $z$- and $y$-components of the vorticity in the $y=0$
plane. Similar to the corresponding panels in Fig.
\ref{fig:jet_flows_prec}, the vorticity can be used to infer the
generation of turbulence and instabilities. Vorticity in both the $y$-
and $z$-directions is produced within the jet itself, as in the
fiducial kinetic jet run. The major difference observed in the runs
with substructures is the increase in the pre-existing ICM vorticity,
which is larger by a factor of $\sim 10$--$100$. Further, any jet-induced
vorticity in the cocoon is subdominant with respect to the level of
vorticity induced by substructures, except possibly in the bottom
left-hand panel, i.e. the run with least time for substructure
motions to stir the ICM, where there appears to be a slight elevation in $\omega_{\rm
  y}$ in the vicinity of the jet cocoon. This agrees with the
behaviour observed in the middle panels, where the pre-existing ICM
gas flows infiltrate and even disrupt the jet cocoon. 

\subsection{Evolution of cluster properties}
\label{sec:evo_halo_sub}

The next point to consider is the impact that the substructures have
on the jets ability to impact the ICM properties. Fig. \ref{fig:substruct_halo_props} shows the radial density (top),
temperature (middle panel) and entropy (bottom) profiles for the
fiducial kinetic jet run (left-hand panel) followed by the haloes that
had undergone $1.25$, $2.25$ and $3.25$ Gyr of stirring by substructure motions
when the jet was initiated, from left to right, respectively. Similar
to Fig. \ref{fig:norm_halo_props}, we show how the profiles change
over time and also include the initial profiles as shown by the dashed
black lines. The first thing to note is that the substructure motions have slightly altered the central gas
  densities and temperatures compared to the idealized cluster.
Further, similar to the injection of the jets in the fiducial runs,
initially the density sharply reduces, while the temperature and
entropy sharply increase in the central regions. However, as also seen
in the earlier runs, cold ICM material, initially displaced by the
expansion of the jet cocoon, flows into the gravitational potential
well of the halo and the profiles return to and somewhat exceed their
initial values.

Specifically comparing to the fiducial kinetic run, it is interesting
to note that the sharp peaks and troughs, at small radii, in the
density, and temperature and entropy profiles, respectively, are not
seen in the runs that have undergone $\sim 2.25$ and $3.35$ Gyr of
stirring by substructures, which instead show rather smoother
profiles. The radial extent to which the AGN heats gas is somewhat
larger ($\sim$few kpc) in these runs compared to the fiducial kinetic jet run. We
attribute these results to the additional pressure support provided by turbulent motions as well as the ability of the substructures' motions
to disrupt the jet cocoon and mix cocoon material with the ICM. 

\begin{figure}
\psfig{file=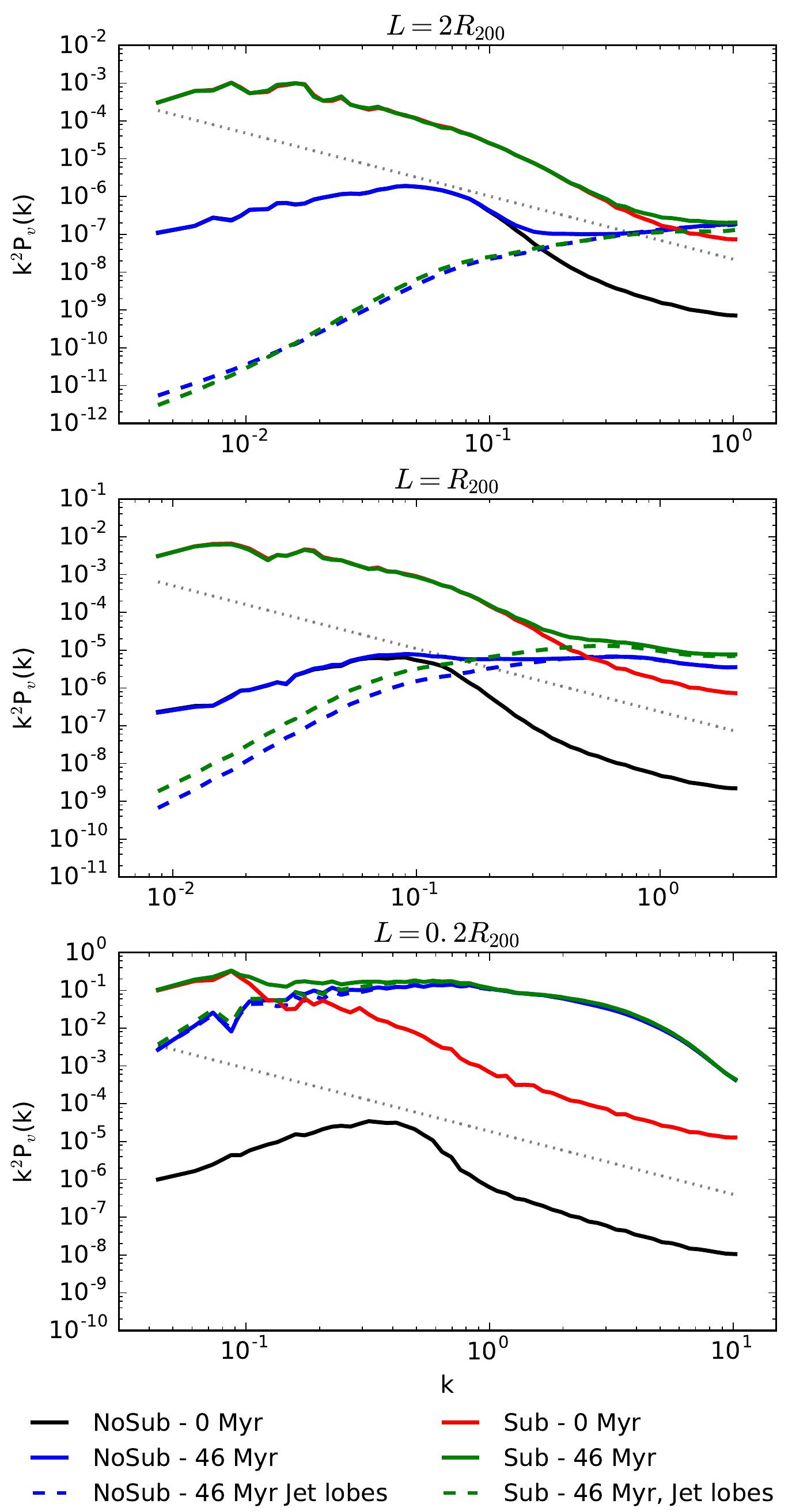,width=0.49\textwidth,angle=0}
\caption{Overview: velocity power spectrum pre- and post-jet inflation for a kinetic jet in a medium that has and has not been stirred by the motions of substructures. In both cases, after the jet inflation, the power is dominated by jet material on scales smaller than the jet length. On larger scales no additional power is seen in the run without stirring, while for the run with stirring the power is dominated by the turbulence driven by the substructures motion.  Coloured dashed curves indicate the power spectra when we only consider jet lobe material, while the grey dashed curve shows the slope of the Kolmogorov turbulence power law of $-5/3$.{Top panel:} power spectra calculated in a box of size $2R_{200}$. {Middle panel:} power spectra calculated in a box of size $R_{200}$. {Bottom row:} power spectra calculated in a box of size $0.2R_{200}$.}
\label{fig:power_spec}
\end{figure}

\subsection{Substructure versus jet-induced turbulence}
\label{sec:substruct_vs_jet_turbulence}

Given that we have introduced subhaloes to stir the ICM and induce
turbulence, it is instructive to consider the $3$D
velocity power spectrum, $P_{\rm v}(k)=|\tilde{\bf v}(k)|^{2}$, where
$\tilde{\bf v}(k)$ is the velocity field in Fourier space. The
simulated velocity field is interpolated on to a regular grid and then
following other similar procedures in the literature
  \citep[e.g.][]{FederrathEtAl09, VazzaEtAl09, Valdarnini2011, BauerSpringel2012, GrisdaleEtAl17}, an FFT algorithm with zero padding is used to
find $\tilde{\bf v}(k)$. $P_{\rm v}(k)$ is plotted in Fig. \ref{fig:power_spec}, where the top, middle and bottom panels show $k^{2}P_{\rm v}(k)$ calculated in boxes of $2$, $1$ and $0.2\times
R_{200}$ on a side, respectively, centred on the BH. For
  all box sizes, we perform kernel-weighted interpolation over $N_{\rm
    ngb}=100$ nearest neighbours on to a $256^{3}$ grid. The power spectra on
  small scales (large $k$) can be sensitive to the  number of
  neighbours used, although we found that the spectra are reasonably
  well converged, provided $N_{\rm ngb}$ is sufficiently large. The adaptive nature of {\sc arepo} and the wide dynamical range covered in
  our simulations (especially after jet activity) unavoidably results in regions
  that are over or under sampled by our chosen grid size. This can
  have a minor effect upon the measured small-scale power \citep[see
    e.g.][for further discussion]{KitsionasEtAl09, Valdarnini2011},
  although does not impact any of our conclusions. This wide dynamic
  range is the primary reason for examining the power in different box
  sizes, while also allowing us to highlight the scales on which the
  jet dominates the velocity power spectrum and its relative
  importance to the global turbulent energy. To this end, the plot
shows $k^{2}P_{\rm v}(k)$ initially and after $t\simeq 45$ Myr for the
fiducial kinetic jet run (black and blue curves, respectively) and for
the run which has undergone $3.25$ Gyr of substructure stirring
(green and red curves, respectively). Further, the dashed curves show
the contribution of jet material, defined as being cells with a jet
mass fraction of $f_{\rm J} > 0.01$, to the total $k^{2}P_{\rm v}(k)$.

First, {we compare} the initial power spectra before the jet has been
inflated for the fiducial kinetic (black) and substructure (red)
runs. It is clear that there is significantly more power within the
run with substructures, for which $k^{2}P_{\rm v}(k)$ roughly follows
the Kolmogorov turbulent power spectrum with a slope of $-5/3$. This
is not surprising given that the motions of the subhaloes stir the ICM,
generate vorticity (see Fig. \ref{fig:substruct_rho_proj}) and drive
turbulence. The next question is whether or not the jet can drive
further turbulence in this system and on what scales? Starting with a
box encompassing the central $(0.2R_{200})^{3}$ region of the cluster,
shown in the bottom panel, we see that after $45$ Myr the power
spectra follow each other very closely and sit above the initial
pre-jet power spectra for large $k$ (small physical scales, $\simlt
30-60$ kpc) and approach similar powers as the initial pre-jet
substructure run only for smaller $k$ values (larger physical scales,
$\simgt 60$ kpc). It is also interesting to note that if we consider
only jet lobe material (dashed curves), the power spectra follow the
corresponding total power spectra almost exactly, indicating that the
power in this box is dominated by the jet. Although we also
  highlight that this does not mean that the lobes dominate the
  kinetic energy, to which they contribute roughly a quarter of the
  budget. However, given the high vorticity (see Section
  \ref{sec:jet_inflation_and_gas_flows}) and velocity dispersion (see
  below) within the lobes, we do expect them to be the primary location
  in which the jets are able to drive turbulence.

Next, we consider a larger box, of volume $r_{200}^{3}$, shown in the
middle panel. {This box} could fit three jets, end to end, along the box
length. In this panel,  we see that the jets only dominate their
corresponding power spectra for large $k$ values (on scales $\simlt
30-60$ kpc). The jet only run (blue curve) remains rather flat and
tends towards the initial pre-jet power-spectrum (black curve) at
small $k$. On the other hand, for the jet injected into the stirred
ICM, the power spectrum (green curve) continues to increase with
decreasing $k$, matching the pre-jet power spectrum (red curve) for
$k\simlt 0.2$ ($\simgt 30$ kpc), both of which approximately follow
the $-5/3$ power law due to the substructure-induced turbulence. When
considering only the jet lobe material, the dashed curves follow the
same shape, albeit with the substructure run having slightly higher
power than the run with no substructures. This shows that the jets
only make a significant contribution to the velocity
  power spectra on small scales ($k\simgt 0.2$, $r\simlt 30$ kpc) and
become subdominant on larger scales. This is particularly evident for
the gas in the run with substructures, in which the substructures'
motions  drive turbulence and are the dominant
contributor to the velocity power spectra.

Finally, the top panel shows $k^{2}P_{\rm v}(k)$ calculated over a box
with sides of length $2R_{200}$, which is considerably larger than the
size of the jet. We see that the pre- and post-jet power spectra in
the run with substructures follow each other exactly for most of the
$k$ values; that is, on these scales the substructures  drive
  turbulence and dominate the velocity power
  spectra. On the other hand, the pre- and post-jet power spectra in
the run without substructures follow each other exactly only below
$k\simeq 0.1$, indicating that the jet is largely ineffective at
 impacting the velocity power spectra on the scale of
this box. This is also evidenced when considering only the jet lobe
material, which reaches low powers for small $k$ values. In essence, on
large scales the jets have no impact on the velocity
  power spectra and therefore do not drive turbulence on these
scales.

\begin{figure*}
\psfig{file=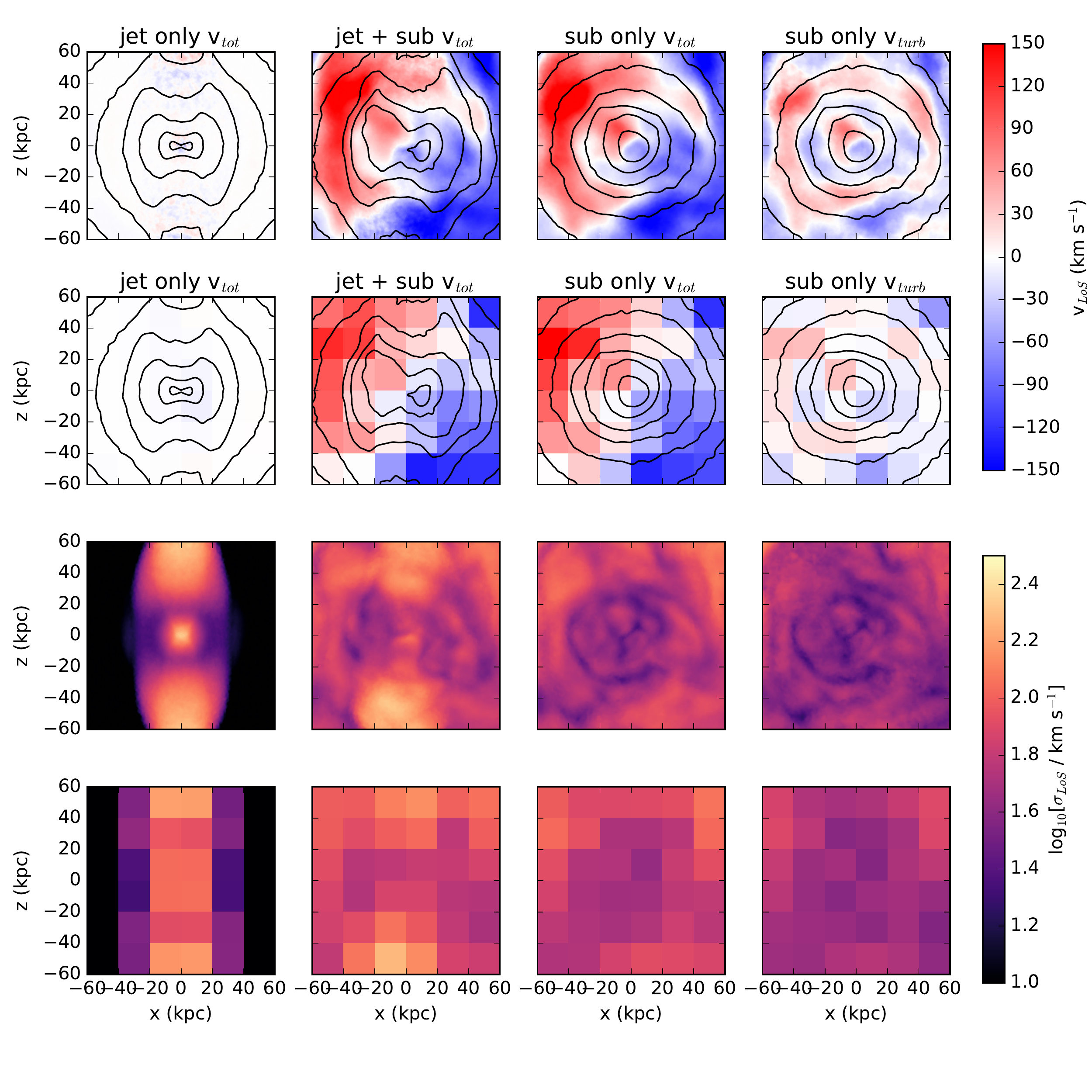,width=0.9\textwidth,angle=0}
\caption{Overview: emission-weighted line-of-sight velocities
  and velocity dispersions within the central $120\times 120$ kpc
  region of simulated clusters, calculated through a depth of $2\times
  R_{200}$ and presented at a resolution comparable to the
  \citet{Hitomi2016} observations of the Perseus cluster and at our
  simulated resolution. Simulations that include both a jet and
  stirring by the motions of substructures can qualitatively reproduce
  the velocity structure and disturbed X-ray emission features
  observed by Hitomi. From left to right, the columns show simulations
  that include just a jet, a jet plus substructures, only motions due
  to substructures and the turbulent velocities due to
  substructures. {Upper panels:} line-of-sight velocities at high
  resolution ({first row}) and Hitomi resolution ({second
    row}), overlaid with X-ray emission contours. {Lower panels}:
  line-of-sight velocity dispersion at high resolution ({third
    row}) and Hitomi resolution ({fourth row}).}
\label{fig:los_v_sig}
\end{figure*}

Additionally, we now consider the emission-weighted line-of-sight
velocities and velocity dispersions within the simulated
clusters. These are shown in the top and bottom panels of Fig.
\ref{fig:los_v_sig}, respectively, both at our simulated resolution
(first and third rows) and $20$ kpc resolution, similar to the Hitomi
satellite for the Perseus cluster (second and fourth rows). From left
to right, the columns show the fiducial kinetic jet run, a kinetic jet
plus substructures and runs only with substructures (no jet) for the
full velocity field {in the third column}, and only the turbulent velocity field in the
fourth column. Turbulent cell velocities are calculated using {a}
multiscale filter method {similar to} \citet{VazzaEtAl12, VazzaEtAl17}. We interpolate the
cell velocities on to a regular grid, with a $3$ kpc resolution \footnote{Note that the measured turbulent velocities can vary systematically with the chosen grid resolution depending on whether the simulated velocity field is over or under sampled. Our chosen value of $3$ kpc $\sim 0.5(n_{\rm cells})^{-1/3}$, where $n_{\rm cells}$ is the number density of cells within the total grid volume, is a compromise to balance between over and under sampling the simulated velocity field \citep[see e.g.][]{Valdarnini2011}.}. The velocity of each grid point can be defined as the sum of the local average velocity and a turbulent component, ${\bf v}_{\rm tot}={\bf v}_{\rm ave} + {\bf v}_{\rm turb}$. For each grid-point the local average velocity,
${\bf v}_{\rm ave}$, is calculated using a
density-weighted average over neighbouring
grid-points. The turbulent velocity is then calculated by subtracting
${\bf v}_{\rm ave}$ from the grid-point velocity. This process is
repeated, calculating ${\bf v}_{\rm ave}$ over an increasing number of
neighbouring grid-points, until {the individual components of} ${\bf v}_{\rm turb}$ {for} a grid-point converge (to within a tolerance factor). This process is performed
for each individual grid-point, providing a uniformly-spaced field of
average velocities. Finally, the local average velocity for each
Voronoi cell is simply taken to be the local average velocity of the
nearest grid-point on which {the} multiscale filtering process was
performed. Subsequently, the turbulent velocity of a Voronoi cell is
estimated by subtracting the local average from the cell's total
velocity. Finally, we include X-ray emission contours on the plots in
the top row, where the X-ray luminosity is calculated using a simple
Bremsstrahlung approximation \citep[e.g.][]{SijackiSpringel06}.

\begin{figure*}
\psfig{file=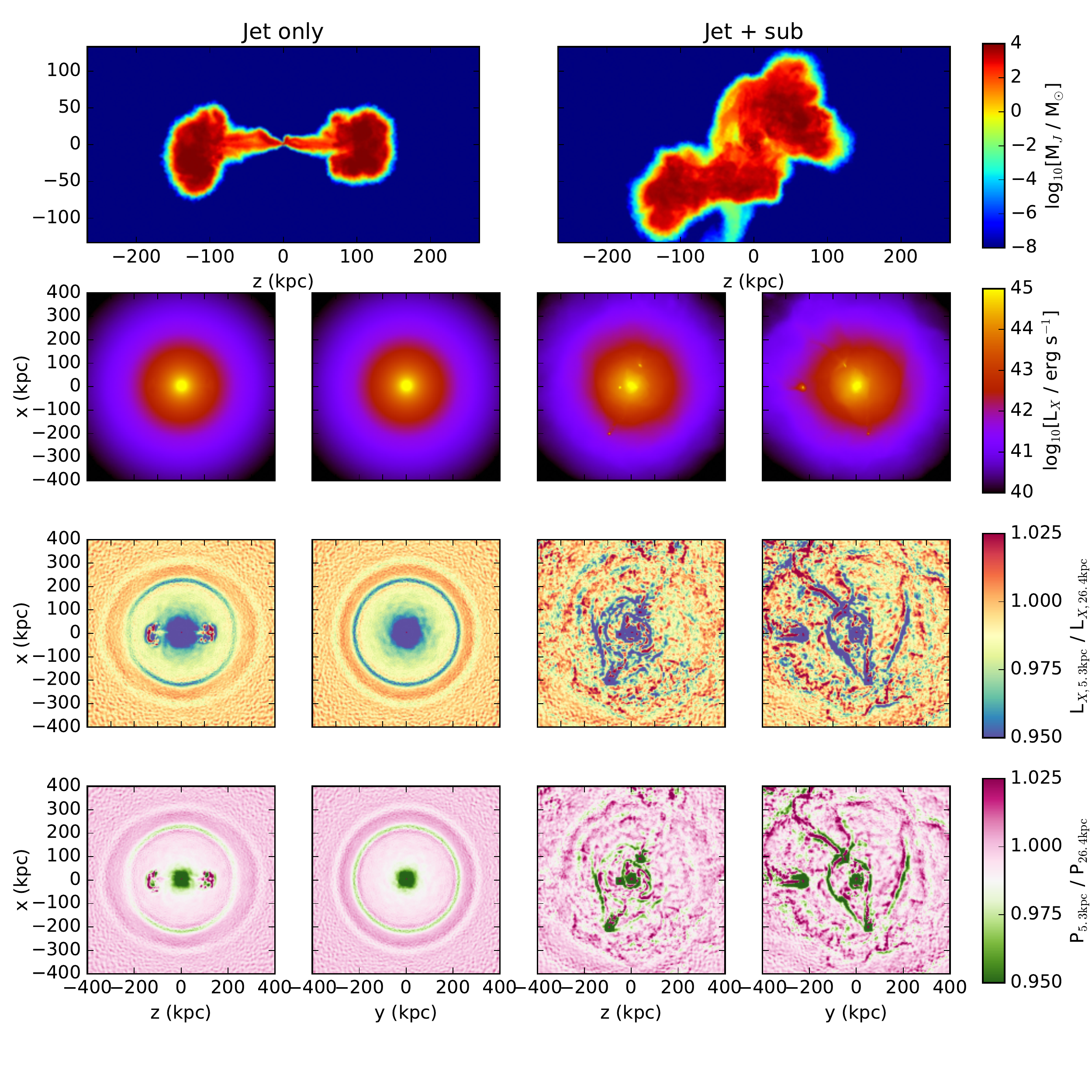,width=0.95\textwidth,angle=0}
\caption{Overview: projected cocoon and ICM maps in the central
  $800$ kpc$^3$ region of a galaxy cluster, $\sim 450$ Myr after jet
  activity has ceased. The two left most columns show a run with only
  a kinetic jet that is switched off after $\sim 20$ Myr, while the
  two right most columns show a run with substructure included. Once
  the jet has switched off, it is possible to observe the propagation
  of sound (pressure) waves into the ICM. Top row: mass of jet
  material along the line of sight, while the jet cocoon remains
  axisymmetric when substructures are not included, this is clearly
  not the case with substructures, which can displace and disrupt the
  relic cavities. {Second row:} X-ray luminosity from the central
  regions. The profiles provide little indication to the existence of
  relic cavities. {Third row:} smoothed X-ray luminosity ratio maps,
  smoothed on different scales. Such maps highlight the positions of jet lobes as well as depressions and enhancements
  in the gas luminosity produced by the propagation of sound
  (pressure) waves. {Fourth row:} smooth pressure ratio maps, similar
  to the third row except showing features in the ICM pressure.}
\label{fig:sound_waves}
\end{figure*}

For the fiducial kinetic jet run, we see negligible line-of-sight
velocities due to jet symmetry, which in this case runs along the
$z$-axis and hence most of the line-of-sight velocities cancel
out. {We note that while viewing the jet at different angles can result in non-zero line-of-sight velocities, they do not exceed $\pm\sim 30$ km s$^{-1}$ at Hitomi resolution.} With regard to the velocity dispersion, we see that this is
dominated by the jet structure, which is in line with the
  jet only being able to drive significant turbulence and hence
  produce the observed velocity dispersion within the jet lobes (see
  discussion above and in Section
  \ref{sec:jet_inflation_and_gas_flows}). At face value, these
results are somewhat inconsistent with the \citet{Hitomi2016}
observations of the Perseus cluster, which show a large shear velocity
across the observed region and a pretty uniform, albeit low, velocity
dispersion across the central region. {However, if} we consider our
runs with substructures (which were not tuned to reproduce specific kinematic properties), both with and without a jet, we see similar line-of-sight velocities, between $\pm\sim 150$ km s$^{-1}$ with a
shear produced across the positive diagonal by motions of
substructures within the central cluster region. This line-of-sight
velocity is dominated by the large-scale bulk motions, given that the
line-of-sight velocity produced by the turbulent gas motions is
significantly smaller (by a factor of 2 or more). Further, when a jet is included the substructure simulations can produce low line-of-sight velocity dispersions consistent with those observed by Hitomi in
the Perseus cluster. In order to produce the distorted features observed in the X-ray luminosity contours, it is necessary to invoke both a jet, which produces the depressions in the X-ray luminosity (compare columns 1 and 3), and substructure motions, which reduce symmetry in the emission (compare
  columns 1 and 2). Finally, we note that while the snapshots used in this figure were chosen for direct comparison with Fig. \ref{fig:power_spec}, we find that the line-of-sight kinematics found in simulations with shorter stirring times (i.e. $\sim 1.25$ and $2.25$ Gyr), and for younger jets can also be consistent with the Hitomi observations with maximum velocity dispersions of $\sigma_{\rm los}\simlt 200$ km s$^{-1}$. This suggests that the gas motions observed by Hitomi within the Perseus cluster core \citep{Hitomi2016} {are consistent with being} driven by a combination of substructure motions and jet activity. We note that {while our simulations only consider a single jet episode, and thus cannot definitively rule out the large-scale kinematics observed by Hitomi being driven by multiple jet outbursts, we note that we are unable to produce a large-velocity shear without including substructure motions, even if we rotate the jet along different sight lines. Further, }our findings are consistent with the recent simulations of \citet{LauEtAl17}, who found that in order to reproduce the kinematic features observed by the Hitomi satellite, a combination of cosmic accretion and jet feedback is necessary.
  
\subsection{Energy transport when the jets are inactive}
\label{sec:sound_waves}

 Up until this point we have only considered a jet which
  is constantly active however, it is instructive to consider the
  evolution of relic cavities and the ICM once a jet becomes
  inactive. We have performed two additional simulations in which the
  jet was switched off after $\sim 20$ Myr and the subsequent
  evolution of system was followed for a further $\sim 450$
  Myr. These simulations have provided two main insights beyond those
  gleaned from our original set of simulations; first, we can
  investigate how substructure motions and the velocity field they
  produce can impact upon relic cavities, and secondly, we are able to
  investigate the propagation of sound and gravity waves within the
  ICM. The top row of Fig. \ref{fig:sound_waves} shows the mass of
  jet material along the line of sight, for runs with a kinetic jet
  with (right-hand panel) and without (left-hand panel) substructures,
  $\sim 450$ Myr after the jet has been switched off. When
  substructures are not included, the jet lobes continue to buoyantly
  rise through the ICM on a path dictated by the original jet
  direction. However, the same fate does not befall the relic cavities
  in the ICM stirred by substructures. While the cavities continue to
  buoyantly rise, they can be significantly displaced from their
  original trajectory, making deductions regarding original jet
  direction difficult. A further interesting point is that stirring by
  substructures can increase mixing of the jet material with the ICM
  by a factor of up to $\sim 3$--$4$, when comparing the ratio 
  \begin{equation}
  f_{\rm mix}=\left(\sum_{f_{\rm J} > f_{\rm cut-off}}{m_{\rm cell}}\right)\bigg/\left(\sum_{{f_{\rm J} > f_{\rm cut-off}}}{f_{\rm J}m_{\rm cell}}\right),
  \end{equation}
for runs with and without substructures, provided that $f_{\rm cut-off} \simlt10^{-5}$. For higher $f_{\rm cut-off}$, values of $f_{\rm mix}$ become similar between the runs, although we note that in both runs no gas exists with $f_{\rm J} \simgt 3\times 10^{-3}$. Further, we find that while $\sim 89$ per cent of jet material still resides within cells with $f_{\rm J} > 10^{-4}$ for the run without substructures, this drops to $\sim 59$ per cent when substructures are included. 
  
Both pressure waves (more commonly referred to as sound waves in the
literature) and gravity waves are expected to be driven by jet
activity and the inflation of jet cocoons
\citep[e.g.][]{OmmaEtAl04, RuszkowskiEtAl04, SijackiEtAl06a,
  SijackiEtAl06b, SijackiEtAl07, ReynoldsEtAl2015, GuoEtAl17}. Observations
appear to indicate the presence of AGN-driven sound waves within the
ICM \citep[e.g.][]{FabianEtAl03, FabianEtAl05b, FabianEtAl17},
although the location at which they deposit their energy intricately
depends upon the ICM viscosity \citep[e.g.][]{RuszkowskiEtAl04,
  SijackiEtAl06b}. On the other hand, while gravity waves have been
seen in simulations of jet feedback \citep[e.g.][]{OmmaEtAl04,
  ReynoldsEtAl2015}, it is expected that they do not carry sufficient
energy to significantly impact ICM cooling \citep{ReynoldsEtAl2015,
  FabianEtAl17}. In the simulations we present here, the expansion of
the cocoon is initially faster than the ICM sound speed and hence
drives a shock into the ICM. However, as the expansion rate of the
cocoon material slows, the shock wave broadens into a sound wave that
propagates into the ICM  {\citep[see also][]{GuoEtAl17}}. The sound wave is
most readily detectable once it has had time to propagate to large
radii and after the jet has become inactive. The location and shape of
this sound wave $\sim 450$ Myr after the jet has been switched off
can be seen in the bottom two rows of Fig. \ref{fig:sound_waves}. The ratio of projected X-ray luminosity (third
row) and pressure (fourth row) maps, smoothed on a scale of $\sim 5.3$
kpc to reduce noise and then divided by maps smoothed on a larger
scale of $\sim 26.4$ kpc, is shown to highlight the existence of
sound waves as enhancements and depressions in the corresponding
maps. For the run without substructures (first and second columns), the
almost circular shape of the sound waves is clearly seen. It is also
evident that as the buoyant rise of the lobes slows below $c_{\rm s,
  ICM}$, the bow shock transitions into a sound wave and detaches
itself from the motion of the lobes. On the other hand, sound waves
and weak shocks produced by substructure stirring completely disrupt
the cocoon in the run with substructures (third and fourth
columns), and hence, we do not {readily detect} the large sound wave in this run. We note that we do not include a description for physical viscosity in our simulations, and so any dissipation of, and hence heating due to, the sound waves is purely numerical.

Finally, as the jet and sound wave propagate through the ICM, they
perturb the ICM, which after their passage can continue to oscillate. \cite{OmmaEtAl04} classified similar oscillations in their simulations, as g-modes excited within the ICM. In our case, this would be consistent with the generation of vorticity perpendicular to the jet motion discussed
in Section \ref{sec:jet_inflation_and_gas_flows}. The gravity waves are
confined to the region within the expanding sound wave and have a very
small amplitude. While a full analysis of their energetics is beyond
the scope of this paper, it is expected that they make an
insignificant contribution to ICM heating \citep{ReynoldsEtAl2015}. 

\section{Discussion}
\label{sec:discussion}

\subsection{Numerical Jets}
\label{sec:discussion_numerical_jets}

In Section \ref{sec:fiducial_runs}, we compared the impact of various
numerical parameters and different energy injection techniques on jet
evolution. We compared the evolution of jet length with simple
analytical predictions for jet evolution, based on those of
\citet{Begelman89}. We found that the jet behaves as expected based on
this simple model, but only when the evolution of the jet
cross-section and momentum rate are taken into account
\citep[e.g.][]{NormanEtAl82, LindEtAl89, KrauseCamenzind01,
  Krause03}. However, to successfully model the propagation and
evolution of a jet, one must make careful consideration to the
refinement scheme implemented within the simulation. We have shown
that an overly aggressive de-refinement scheme can stunt jet
evolution, inhibit growth and promote numerically driven mixing.

A number of energy injection schemes are used within the
literature. Earlier works \citep[e.g.][]{
, CattaneoEtAl07}
injected momentum and thermal energy, while more recent work
\citep[e.g.][]{GaspariEtAl11, PrasadEtAl15, YangReynolds16a,
  YangReynolds16b} implement purely kinetic jets, thus motivating us
to test different schemes. While the thermal jets may seem more
physically motivated, given that momentum is intrinsically conserved,
they can significantly alter the entropy profiles of galaxy clusters {and readily destroy cool cores
\citep{CattaneoEtAl07}. On the other hand, it has been shown in the literature that purely kinetic jets are able to maintain cluster cool cores \citep{GaspariEtAl11, YangEtAl2012}}. We find that although the majority of the jet kinetic energy is thermalized through shocks within the jet lobe, irrespective of the
chosen energy injection technique, there are still differences in jet
morphology. While the purely kinetic injection produces longer jets,
more akin to an FR-II morphology, the thermal and precessing kinetic
jets appear much closer to the FR-I jets seen in galaxy clusters.

The requirement for jets to precess in order to efficiently and
isotropically heat the ICM was proposed by \citet{VernaleoReynolds06},
and has since been implemented in a number of works as a necessary
ingredient \citep[e.g.][]{Falceta-GoncalvesEtAl10,
  LiBryan14,YangReynolds16a,YangReynolds16b}. The precessing jet we
consider is the same as the kinetic jet except that the direction of
the jet precesses about the $z$-axis at an angle of $15^{\circ}$ with
a period of $\sim 10$ Myr, similar to the implementation of
\citet{YangReynolds16a}. Morphologically, the precessing jet appears
similar to the thermal jet, with a seemingly more isotropic jet lobe
distribution. In terms of energetics, a larger fraction of the initial
kinetic energy of the precessing jet is converted into thermal energy
through shocks due to the extra jet motions. Despite the
  morphological differences, as discussed in Section
  \ref{sec:evo_cluster_props}, both the fiducial kinetic and
  precessing kinetic jets produce rather similar radial ICM profiles,
  with the precessing jet only being moderately better at preventing cold
  material reaching the central regions of the halo. {However, we note that \citet{MeeceEtAl17} have recently compared different precession angles, finding that larger values of $\theta_{\rm prec}$ result in a larger fraction of jet kinetic energy thermalizing through shocks.}

 Observations of the locations of {\it relic} X-ray lobes suggest
 that AGN jets may be able to move, while the shape of observed jet
 emission further suggests precession \citep[e.g.][]{DunnEtAl06,
   Marti-VidalEtAl11, BabulEtAl13}. Although as shown in
   Section \ref{sec:sound_waves}, lobes can also be displaced by ICM bulk
   motions. More recent observations of the jet and molecular outflow
 in NGC 1377 \citep{AaltoEtAl16} show kinematic behaviour that is
 also consistent with jet precession, as is also found in the line-of-sight velocities of our
 precessing jet model. The precession of jets produced by SMBHs is
 still debated in the literature, with the exact mechanism driving the
 precession not clearly understood. \citet{NixonKing13} have
 considered whether or not jets are physically able to precess or even
 move based upon either the evolution of the orientation of the BH
 spin or inner accretion disc angular momentum. They suggest that for
 massive BHs, such as those in AGN, it is very difficult to
 significantly change the BH spin direction during a single accretion
 event and that jet precession time-scales of less than the accretion
 time-scale would imply that it is the accretion disc driving the jet,
 opposed to the spin. In this case the jet precession could
 potentially be driven by self-induced warping of an irradiated
 accretion disc \citep{Pringle96, Pringle97} or by accretion discs
 that tear \citep{NixonEtAl12a, NixonEtAl12b} due to the
 Lense-Thirring effect \citep{LenseThirring1918}. On the other hand,
 alternative models also suggest that massive BH binaries could also
 result in jet precession \citep{BegelmanEtAl80}.
   
\subsection{Jet inflation and velocity structure}
\label{sec:disc_jet_inflation}

The jet lobe inflation and cocoon development proceed through the interaction of the jet
with the ambient ICM. Somewhat akin to AGN-driven winds \citep{FQ12a,
  ZK12a}, the jets collide with and shock against the ICM. {This produces}
the jet lobes, full of shocked jet material, and the rest of the jet
cocoon. Initially, while the cocoon material expands faster than
 $c_{\rm s, ICM}$, the cocoon consists of shocked ICM
  material. However, once the expansion slows, the shock broadens into
  a sound wave expanding into the ICM. The
general structure of these regions is illustrated in Fig.
\ref{fig:jet_structure}, which shows the four main regions in the
vicinity of a jet. The shocked jet material expands
thermally, producing the jet lobes, while the cocoon
  expands perpendicular to the jet direction either as a shock wave or
  a sound wave, depending on the gas radial velocity. While the propagation of the jet in the
$z$-direction drives a bow shock into the ICM, which is dominated by
ram pressure, the shocked jet and ICM material is dominated by thermal
pressure. This is in line with the theoretical structure of jets
outlined by analytical models \citep[e.g.][]{BlandfordRees1974, Scheuer1974, Begelman89} and simulations
\citep[e.g.][]{NormanEtAl82, LindEtAl89, KrauseCamenzind01, Krause03,
  HardcastleKrause13}.

As discussed in Section \ref{sec:jet_inflation_and_gas_flows}, at later
times, ICM material displaced by the inflation of the jet 
  cocoon flows into the centre of the galaxy cluster, mainly through
the plane perpendicular to the jet direction. Some of this material
then appears to be dragged up by the motion of the jet before falling
back into the cluster potential well, resulting in a gas circulation
towards the base of the jet, similar in fashion to a galactic
fountain. We also note that backflows, {expected to arise due to steep gradients in entropy and density \citep[e.g.][]{ADS2010, CieloEtAl2014}, form within the jet cocoon. However, especially at later times, we find that these are readily disrupted by jet-driven turbulence.}

\subsection{Heating mechanisms and the energy budget}

 The combined thermal and kinetic energy content of the hot-jet-enriched lobe material ($f_{\rm J} > 0.01$) makes up $\sim 30$ per cent of the total
injected energy and is dominated by the thermal component. We find that significant vorticity, and hence
turbulence, is only produced within the jet lobe material \citep[see
  also][]{WeinbergerEtAl17} and given that most of the energy content
of the jet lobe is thermalized through shocks, we suggest that the jet is unable to
drive significant turbulence in the ICM. After $45$ Myr, the remaining $\sim
70$ per cent of energy injected by the jet, which does not reside in the jet-enriched lobe gas, either goes in to less jet-rich lobe gas ($0.001 < f_{\rm J} < 0.01$, $\sim 10$ per cent), the kinetic energy of the
expanding cocoon or other kinetic motions, the gravitational potential energy of gas lifted out of the potential well of the cluster, heating of the ICM (see below) or is lost to
radiative cooling and adiabatic expansion. {Considering the kinetic run {\it without} radiative cooling (see Fig. \ref{fig:jet_energy_budget}), $\simeq 64$ per cent of the injected energy goes into the thermal component, with about half of this being in ICM gas with $f_{\rm J} < 0.001$. Given that the jet is unable to drive significant turbulence within the ICM and that we do not find significant mixing of jet lobe and ICM material, we suggest that this ICM heating, at least in the adiabatic case, is primarily due to compression and weak shocks \citep[see also][]{YangReynolds16b}.} We note that this partitioning of the energy is achieved while the jet is active and will likely change as the system evolves once a jet becomes inactive.

As outlined in the Introduction section, a number of channels have been
proposed for converting the kinetic energy of the jet into thermal
energy within the ICM to suppress cooling. The most direct interaction
between the jet and the ICM is through shocks. While we
  find that a significant fraction of the kinetic energy of the jet
is thermalized within the jet lobes themselves, the expansion of the cocoon is significantly less
  effective at driving {strong} shocks into the ICM. Although a continuous bow
  shock is driven in the jet direction, the perpendicular expansion of
  the cocoon only drives a shock wave into the ICM during the first
  few Myr of jet activity. This then transitions into a sound wave
  propagating into the ICM (see Section \ref{sec:sound_waves}). The lack of
  strong ICM shocks in our simulations is consistent with the fact
  that observed AGN-driven shocks in galaxy clusters are often
  weak \citep[e.g.][]{FabianEtAl06, FormanEtAl07, CrostonEtAl11,
  SandersEtAl2016} and are not expected to provide enough energy to be the
dominant contribution to the heating of the ICM, although may be able to prevent cooling close to the BH
\citep{NulsenEtAl07}. Indeed, \citet{YangReynolds16a} found that while
weak shocks can heat the ICM, they cannot overcome radiative cooling
and only result in a {`reduced cooling flow'}, while
\citet{LiEtAl2016} found that shock heating can provide an order of
magnitude greater heating than turbulent heating. It is
  also interesting to note that, as shown in Section \ref{sec:sound_waves},
  once the jet becomes inactive and the buoyantly rising bubbles slow down,
  the sound wave produced by the jet inflation `detaches' from the
  jet lobes and can propagate to large distances through the
  ICM  {\citep[see also][]{GuoEtAl17}}. Similarly, sound waves have been observed in galaxy clusters
  \citep[e.g.][]{FabianEtAl03, FabianEtAl05b}, but the exact details
  of how and where the energy carried by these sound waves is
  dissipated depends upon the form of the  physical viscosity of the
  ICM \citep[e.g.][]{RuszkowskiEtAl04, SijackiEtAl06b}.

It has been suggested that a significant amount of energy can go into the form of cavity heating \citep[e.g.][]{ChurazovEtAl02, BirzanEtAl04,NulsenEtAl07}, whereby the potential energy of the material displaced by the
expansion of the jet lobes can be converted into kinetic energy and
subsequently heat. As highlighted in Section
\ref{sec:jet_inflation_and_gas_flows}, the jet lobes
inflated in our simulations are able to displace large quantities of
ambient ICM gas ({$\sim10^{10}-10^{11}$ M$_{\odot}$ by $45$ Myr, depending on energy injection method}), some of which likely falls into the potential well of the cluster. Thus the jet action not only stimulates the
conversion of ICM gravitationally potential energy into kinetic energy
and heat but also provides further fuel for BH growth. 

Finally, mixing of jet material could play a role in
communicating the thermalized jet energy within the lobes to the
ICM. Similarly to \citet{YangReynolds16a}, we find that there can be
mixing of jet material within the jet lobes, but find negligible amounts of jet material beyond their immediate
  vicinity while the jet is active. The evolution of the jet lobes
  once the jet switches off depends upon whether or not substructures
  have stirred the ICM. In the run without substructures, the jet
  lobes rise buoyantly through the ICM and are gradually disrupted by
  fluid instabilities. Additionally, the motions of substructures
  displace the rising lobes from their original trajectory and further
  promote  mixing. We find that $\sim 450$ Myr after the jet switched
  off, substructures can promote mixing by a factor of up to $\sim 3-4$.
Interestingly, \citet{HillelSoker16b, HillelSoker17} suggest from their simulations
that mixing of bubble material with the ICM actually plays a more
dominant role than turbulent and shock heating. However, we note that the draping of magnetic field lines over the jet lobes, even in a weak magnetic field, could have an important impact on the evolution and dynamics of jet lobes \citep[e.g.][]{DursiPfrommer08}. While recent magnetohydrodynamic simulations have found that mixing is less efficient than in purely hydrodynamic simulations \citep[e.g.][]{WeinbergerEtAl17}. {On the other hand, it has also been shown that anisotropic thermal conduction can in fact increase mixing and promote isotropization of injected feedback energy \citep{KannanEtAl17}.}

\subsection{Substructures and turbulence in the ICM}

In Section \ref{sec:sub_structure}, we built upon the idealized simulations
presented in Section \ref{sec:fiducial_runs} by including substructures,
which were added by hand. We found that the motions of the
substructures are able to produce significant vorticity within the ICM
and drive turbulence, with the total kinetic energy of gas accounting
for $\sim 9$--$30$ per cent of the total {kinetic plus thermal} energy within the cluster virial
radius. We have shown that while jet induced motions are able to drive turbulence and dominate the velocity power spectrum on scales smaller than the jet length, any large-scale
turbulence is likely to be driven by substructure motions within the
cluster. In fact, the line-of-sight velocities and
velocity dispersions measured in our simulations with substructures
are consistent with those measured by Hitomi observations of the
Perseus cluster \citep{Hitomi2016}, suggesting a potentially important
role for substructure-driven turbulence in the energy budget of the
ICM. {However, given that a number of different simulations all find levels of turbulence consistent with the Hitomi results \citep[e.g.][]{ReynoldsEtAl2015, YangReynolds16a, HillelSoker16, LauEtAl17, WeinbergerEtAl17}, but reach at different conclusions regarding what heating mechanisms are dominant, we believe that it is not possible to determine the main contribution to the energy budget from Hitomi observations alone.} 

Observations suggest that turbulence exists within the ICM
\citep{SandersEtAl2010,SandersEtAl2011,
  ZhuravlevaEtAl12, ZhuravlevaEtAl2014, SandersFabian2013, PintoEtAl2015, Hitomi2016,
  OgorzalekEtAl17}, perhaps contributing $\simlt 4$--$40$ per cent of the
pressure support within the ICM \citep[e.g.][]{SandersFabian2013,
  PintoEtAl2015, Hitomi2016}. Such turbulence is likely to be produced
by {\it large scale} processes within the cluster, such as sloshing,
accretion, mergers and substructure motion \citep{DolagEtAl2005,
  VazzaEtAl12, GuEtAl2013, ZuHoneEtAl2013,
  IapichinoEtAl17, VazzaEtAl17}. However, it has also been proposed that jets could
drive turbulence sufficient to offset cooling within galaxy clusters
\citep{BanerjeeSharma2014, ZhuravlevaEtAl2014}. Such turbulence could be produced when AGN feedback excites $g$-modes in the ICM, which then decay into volume-filling turbulence \citep{ReynoldsEtAl2015}.
Analysis presented by \citet{ZhuravlevaEtAl2014} for the Perseus {and Virgo} clusters has shown that the
observed levels of turbulence would be able to balance loses due to
radiative cooling. However, as discussed in \citet{FabianEtAl17}, the propagation velocity of $g$-modes in the Perseus cluster would likely fall an order of magnitude short of that necessary for the dissipation of AGN-driven turbulence to balance radiative cooling. Indeed, when analysing the simulations presented here we
found that while the jet is able to drive turbulence within the
 jet lobes, it is unable to drive large-scale turbulence
within the ICM. The inability to drive significant turbulence within
the ICM is consistent with previous simulations
\citep[e.g.][]{ReynoldsEtAl2015, YangReynolds16a,
  WeinbergerEtAl17}. Therefore, the only scenario in which jet-driven
turbulence could provide an isotropic source of heat is
if the jet lobes fill the cooling radius of the cluster, as in M87 \citep{FormanEtAl07}. However, even in our
  purely kinetic jet runs we find that the {\it total} kinetic energy
  content of the lobes accounts for only $\sim 4$ per cent $E_{\rm Inj}$ and therefore does not dominate the energy budget.

\subsection{Limitations of current simulations}

 Given the wide dynamical range and vast number of
  physical processes that could be important in modelling AGN jets in
  galaxy clusters, there are by necessity a number of limitations to
  the simulations presented here.  The current simulations have used
a fixed $\dot{m}$ accretion rate and hence fixed jet power throughout
and thus do not include the back reaction of the ICM on to the
subsequent accretion rates and jet production. While our set-up has
allowed us to make clean comparisons between different jet injection
techniques, in order to make a more meaningful interpretation of how
jets regulate heating and cooling within the ICM, we will need to
include self-consistent accretion and feedback in future work. On top
of this, the jet direction, or the axis about which the jet precesses,
is fixed in the simulations presented here, while in reality this is expected to be linked to either the
BH spin or accretion disc angular momentum
\citep[e.g.][]{NixonKing13}. We hope to remedy these shortcomings in future work by combining the jet feedback model outlined in this paper with a newly developed BH accretion scheme, which not only
tracks the accretion rate of gas on to the BH but also models the
evolution of both the accretion disc and BH spin (Fiacconi, Sijacki \& Pringle, in preparation).

 This work also focuses on purely hydrodynamic jets
  and does not include the effects of magnetic fields, the importance
of which, for example through inhibiting mixing, has been highlighted in recent simulation works
\citep[e.g.][]{EnglishEtAl16, WeinbergerEtAl17}. However, we find that
it is both instructive and important to understand the hydrodynamic
evolution of the jets prior to adding further physics, while also
presenting a model that can be readily implemented into hydrodynamic
cosmological simulations. A further limitation of the current work is
one which plagues many other large-scale simulations of jet evolution,
in that the hot gas component of the jet lobe is modelled as a
non-relativistic ideal gas. First, beyond temperatures of $T\sim
10^{10}$ K, any electron population will be relativistic, and
secondly, it is not clear what the exact composition of physical jets
and jet lobes is, nor what is the relative importance of leptonic and
hadronic components \citep[e.g.][]{DunnEtAl06b, BirzanEtAl08,
  CrostonEtAl08, CrostonHardcastle14, KangEtAl14, KawakatuEtAl16}.

Finally, while we have attempted to include the effects of the motions
of substructures by introducing them by hand in an idealized system,
we include neither a fully live dark matter distribution nor, perhaps
more importantly, a fully cosmological galaxy cluster evolved
self-consistently as a function of cosmic time. This will potentially
impact the large-scale turbulent velocity field and thus the
subsequent evolution and interaction of the jet, jet cocoon and
ICM. Our goal is to perform such simulations in upcoming work.

\section{Summary}
 \label{sec:summary}

In this paper, we have presented a novel approach for the simulation
of AGN jets in the moving mesh-code {\sc arepo}
\citep{SpringelArepo2010}. The main results of this paper are as
follows:

\begin{itemize}
\item {With an appropriate refinement scheme, we are able to
  successfully model the injection of a hydrodynamic jet and the
  subsequent inflation of jet lobes that are consistent with
  analytical expectations and previous simulation work. Different
  energy injection methods can result in jets with greatly differing
  morphologies; however, the total energy content within jet lobe
  material is remarkably consistent between methods (assuming energy
  is explicitly conserved in the injection process).}

\item {The jets are able to affect the energy budget within the
  central regions of a galaxy cluster, changing the thermal, kinetic
  and gravitational potential energy content. The jets are able to generate significant levels of vorticity and drive turbulence within the jet lobes. However, such turbulence is not seen on larger scales (unless substructures are included) and we suggest that jets are unable to drive
  turbulence within a significant fraction of the ICM \citep[see
    also][]{ReynoldsEtAl2015, YangReynolds16a, WeinbergerEtAl17}.}

\item {Substructures within the galaxy cluster are able to stir the
  ICM and generate turbulent motions. This additional velocity field
  can interact with and disrupt the cocoons inflated by jets, providing additional pressure support,
  potentially promoting mixing of jet cocoon material with the ICM,
  and resulting in less symmetric jets.}

\item {Simulations that include substructure motions and a jet are
  able to produce line-of-sight velocity and velocity dispersion maps,
  and X-ray emission contours, consistent with those observed in the
  Perseus cluster by \citet{Hitomi2016}. Therefore, we conclude that {it is possible to produce} the
  low levels of turbulence that are observed within the Perseus
  cluster {through a combination of stirring of the ICM by substructure
  motions on large scales and jet feedback on smaller scales.}}
  
\end{itemize}

\section*{Acknowledgments} 

We would like the thank the anonymous referee for their constructive report on this manuscript. We would also like to thank Christoph Federrath, Tiago Costa and Noam Soker for helpful comments and suggestions. MAB and DS acknowledge support by the ERC starting grant 638707 `BHs and their host galaxies:
co-evolution across cosmic time'. DS further acknowledges support
from the STFC. This research used: the DiRAC Darwin Supercomputer
hosted by the University of Cambridge High Performance Computing
Service (http://www.hpc.cam.ac.uk/), provided by Dell Inc. using
Strategic Research Infrastructure Funding from the Higher Education
Funding Council for England and funding from the Science and
Technology Facilities Council; the COSMA Data Centric system at Durham
University, operated by the Institute for Computational Cosmology on
behalf of the STFC DiRAC HPC Facility. This equipment was funded by a
BIS National E-infrastructure capital grant ST/K00042X/1, STFC capital
grant ST/K00087X/1, DiRAC Operations grant ST/K003267/1 and Durham
University. The DiRAC Complexity system, operated by the University of
Leicester IT Services, which forms part of the STFC DiRAC HPC Facility
(www.dirac.ac.uk).  This equipment is funded by BIS National
E-Infrastructure capital grant ST/K000373/1 and STFC DiRAC Operations
grant ST/K0003259/1.  DiRAC is part of the UK National
E-Infrastructure.

\bibliographystyle{mnras}

\bibliography{ArepoJet}

\appendix

\section{Jet mass}
\label{App:JetMass}

\begin{figure}
\psfig{file=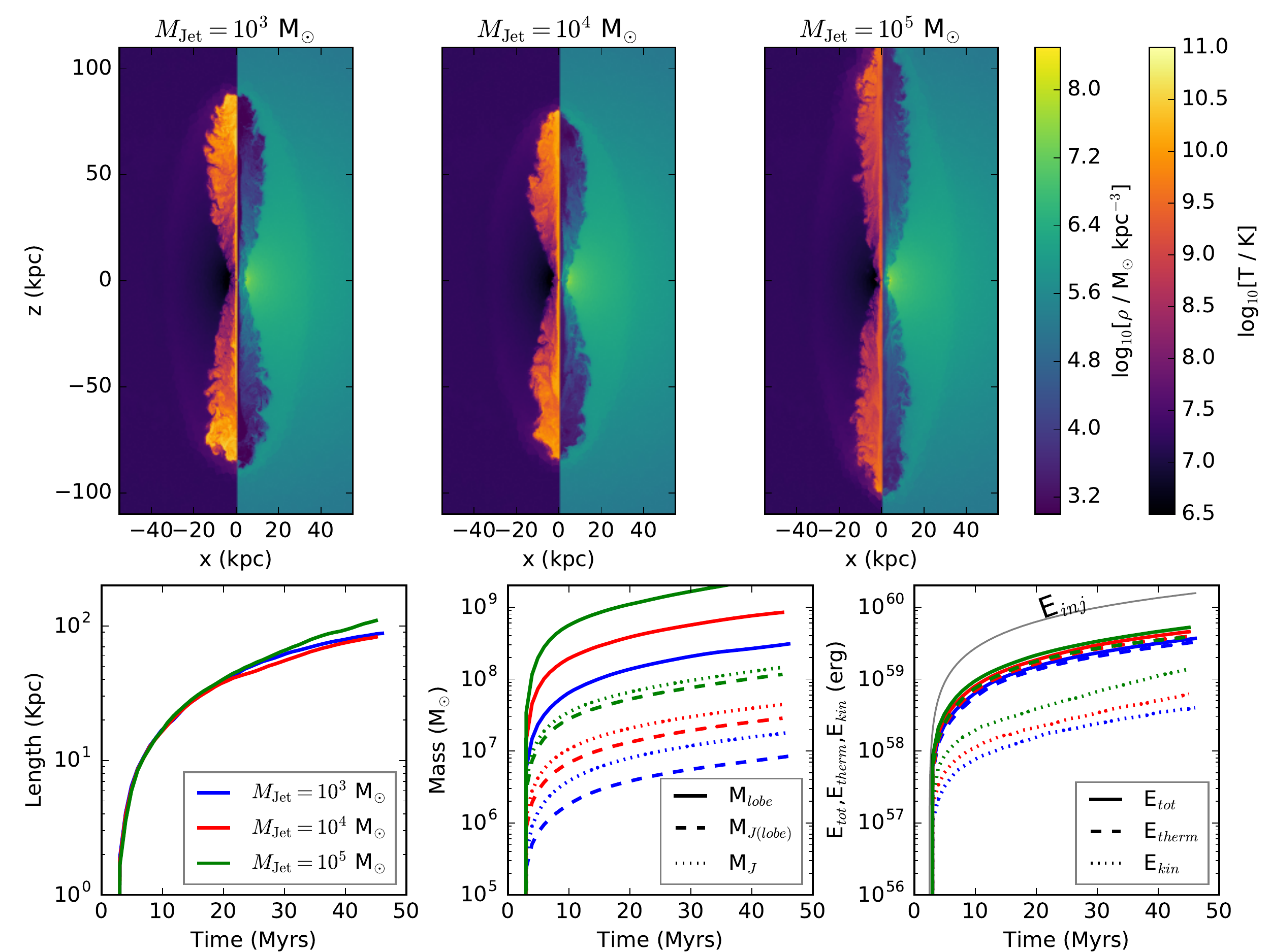,
  width=0.5\textwidth,angle=0}
\caption{Overview: dependence of jet evolution on jet cylinder
  mass, which show remarkably similar jet morphologies, although jet
  masses and kinetic energies correspondingly increase with jet
  cylinder mass. Top row: density and temperature slices through
  the $y=0$ plane at $t\simeq 45$ Myr. Bottom row: evolution of
  the jet length (left-hand panel), different components of jet mass
  (middle panel) and different components of jet energy content
  (right-hand panel) for the kinetic runs with $M_{\rm Jet}=10^{3}$
  (left-hand panel and blue curves), $10^{4}$ (middle panel and red
  curves) and $10^{5}$ M$_{\odot}$ (right-hand panel and green
  curves). For comparison, we show the total injected jet energy
  (equation \ref{eq:e_inj}) by the solid black line in the lower
  right-hand panel.}
\label{fig:jet_overview_jet_mass}
\end{figure}

In our fiducial runs, we haved chosen a jet cylinder mass of $M_{\rm
  Jet}=10^{4}$ M$_{\odot}$, as a balance between resolution and
numerical resources. A large $M_{\rm Jet}$ leads to a poorly resolved
jet, while a small $M_{\rm Jet}$ leads to increased run times. To check
the impact on jet properties, we have performed additional simulations
of kinetic jets with $M_{\rm Jet}=10^{3}$ and $10^{5}$
M$_{\odot}$. Density and temperature slices after $t\simeq 45$ Myr for
these runs are shown in the top left-hand and right-hand panels of Fig.
\ref{fig:jet_overview_jet_mass}, along with the fiducial $M_{\rm
  Jet}=10^{4}$ M$_{\odot}$ run in the top middle panel.  It is clear
from these slices that increased jet masses result in higher jet
densities and lower temperatures, as approximately the same thermal
energy is spread over the larger mass. Additionally, on the bottom row
we show the evolution of jet properties, similarly to previous
figures.

In all runs, the structure and morphology of the jet appears similar,
although the $M_{\rm Jet}=10^{5}$ M$_{\odot}$ jet is slightly longer. As
expected, increasing $M_{\rm Jet}$ leads to more massive jets, as shown
in the bottom middle panel, with the jet lobe mass, total jet mass and
jet mass within the lobe all increasing with $M_{\rm Jet}$. As shown in the lower
right-hand panel, $M_{\rm Jet}$ also impacts the energy content of the
jet, which increases for larger $M_{\rm Jet}$ and is especially evident
for the kinetic energy. If we consider the total energy content within
jet lobe material, we find that the $M_{\rm Jet}=10^{4}$ and $10^{5}$
M$_{\odot}$ jets retain $24$ per cent and $42$ per cent more energy in the lobe
material compared to the $M_{\rm Jet}=10^{3}$ M$_{\odot}$ jet. This
again illustrates the difference in mixing between the jet masses.

\section{Kernel function}
\label{App:Kernel}

\begin{figure}
\psfig{file=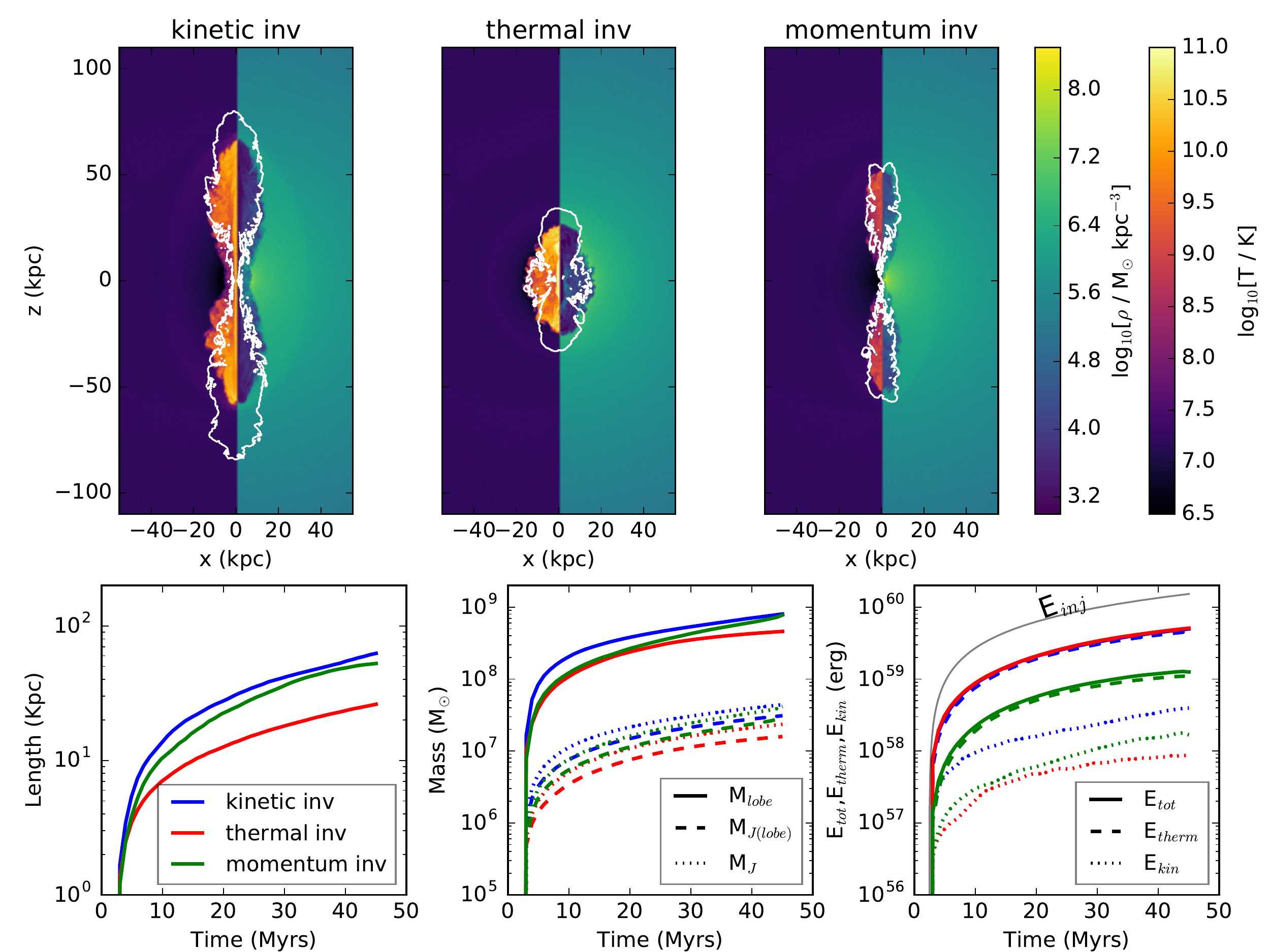,
  width=0.5\textwidth,angle=0}
\caption{Overview: dependence of jet evolution on the kernel
  weighting scheme implemented; general jet properties remain similar
  although inverting the weighting scheme with respect to $z$ can
  result in broader, shorter jets. Top row: density and
  temperature slices through the $y=0$ plane at $t\simeq 45$ Myr. Bottom row: evolution of the jet length (left-hand panel), different components of jet mass (middle panel) and different
  components of jet energy content (right-hand panel) for the kinetic
  (left-hand panel and blue curves), thermal (middle panel and red
  curves) and momentum (right-hand panel and green curves) runs, with
  the modified kernel function. For comparison, we show the total
  injected jet energy (equation \ref{eq:e_inj}) by the solid black
  line in the lower right-hand panel.}
\label{fig:jet_overview_inv}
\end{figure} 

Both physical and numerical considerations need to be taken into
account when choosing a suitable kernel weighting function for mass,
momentum and energy injection. In our fiducial models, we have chosen
a similar kernel weighting scheme as has been used by previous authors
\citep[e.g.][]{OmmaEtAl04, CattaneoEtAl07, YangEtAl2012}.  Here we
consider a modification to the scheme by weighting cells closer to the
BH more heavily. This is done by using a kernel of the form
\begin{equation}
W_{\rm J}(r, z)\propto{\exp\left(-\frac{r^{2}}{2r_{\rm
      Jet}^{2}}\right)(h_{\rm Jet}-|z|)},
\label{eq:kernel_inv}
\end{equation}
which differs from equation (\ref{eq:kernel}) by a factor of $(h_{\rm
  J}-|z|)/|z|$. This results in material close to the BH receiving a
larger kick, reducing the central density and hence resulting in a
larger $r_{\rm Jet}$, when compared to runs with the fiducial kernel
function (equation \ref{eq:kernel}). Qualitatively, the impact of this
can be seen in the top row of Fig. \ref{fig:jet_overview_inv}, which
shows density and temperature slices similar to those presented in
Fig. \ref{fig:jet_overview_norm}. Additional white contours are
included, outlining the shape of the corresponding jets in Fig.
\ref{fig:jet_overview_norm}, for comparison. In general, the structure
of the jet is similar, in each case, to those presented previously,
with the momentum runs (right-hand panel) being almost
identical. However, in the kinetic (left-hand panel) and thermal
(middle panel) jet runs, the jets are shorter. The reduced length of
the jets can be attributed to the increased jet cylinder radius,
$r_{\rm Jet}$, which results in broader jets and hence a larger ram pressure force acting on the
jet along the $z$-axis, as discussed in Section \ref{sec:analytical}. The
bottom row of Fig. \ref{fig:jet_overview_inv} shows jet properties
for the kinetic (blue), thermal (red) and momentum (green) runs with
the alternative kernel function. In agreement with visual appearance,
the overall behaviour of the jets is similar to those presented in
Fig. \ref{fig:jet_overview_norm}, with the momentum runs being
almost identical, while jet lengths in the kinetic and thermal runs
are shorter.

\section{Refinement parameters}
\label{App:JetRef}

\begin{figure}
\psfig{file=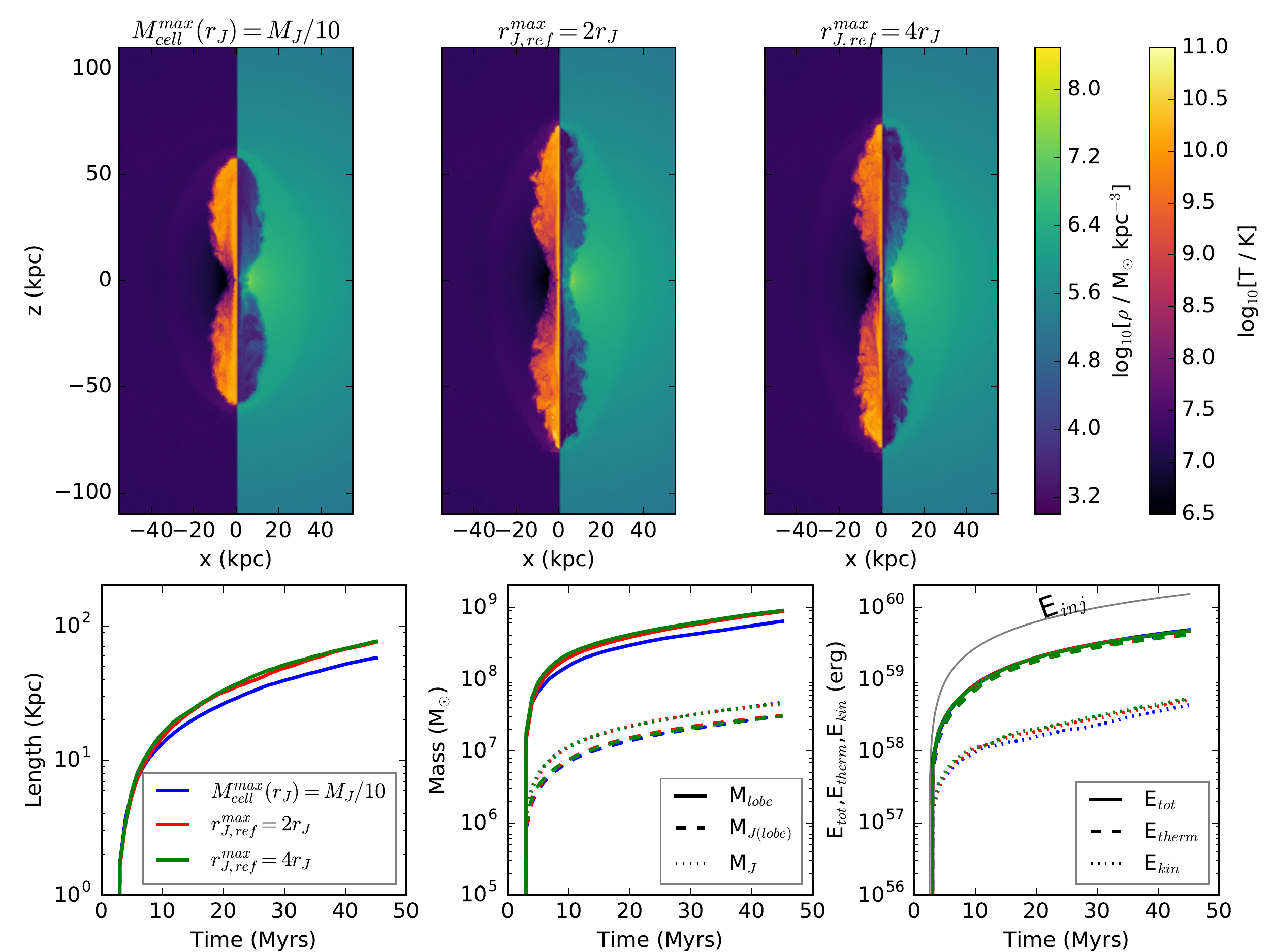,
  width=0.5\textwidth,angle=0}
\caption{Overview: dependence of jet evolution on jet refinement
  parameters. Varying the size of the refinement region has a
  negligible impact on the jet properties; however, increasing the
  maximum cell size can impact the jet morphology. Top row:
  density and temperature slices through the $y=0$ plane at $t\simeq
  45$ Myr. Bottom row: evolution of the jet length (left-hand
  panel), different components of jet mass (middle panel) and
  different components of jet energy content (right-hand panel) when
  $M_{\rm cell}^{\rm max}\left(r_{\rm Jet}\right)=M_{\rm Jet}/10$
  (left-hand panel and blue curves), $r_{\rm Jet, ref}^{\rm
    max}=2r_{\rm Jet}$ (middle panel and red curves) and $r_{\rm Jet,
    ref}^{\rm max}=4r_{\rm Jet}$ (right-hand panel and green
  curves). For comparison, we show the total injected jet energy
  (equation \ref{eq:e_inj}) by the solid black line in the lower
  right-hand panel.}
\label{fig:jet_props_jet_ref}
\end{figure}

The top panels of Fig. \ref{fig:jet_props_jet_ref} show density and
temperature slices for jets after $t\simeq 45$ Myr, while the bottom
panel shows the evolution of the jet length (left-hand panel), jet lobe
mass components (middle panel) and jet lobe energy components
(right-hand panel) when we increase the maximum cell mass at $r=r_{\rm
  Jet}$ to $M_{\rm cell}^{\rm max}\left(r_{\rm Jet}\right)=M_{\rm
  Jet}/10$ (top left-hand panel and blue curves; note that the maximum
cell mass at $r=0$ is also increased by a factor of $10$), decrease the
jet refinement region to $r_{\rm Jet, ref}^{\rm max}=2r_{\rm Jet}$
(top-middle panel and red curves) or increase it to $4r_{\rm Jet}$
(top left-hand panel and green curves). 

While changing the size of the jet refinement region has little impact
on the morphology and physical structure of the jet, allowing larger
cell mass results in a shorter jet, likely due to the effective
courser jet resolution providing a larger jet working surface radius
against the ICM. However, the evolution of the jet lobe mass and, in
particular, jet lobe energy content remain remarkably consistent
between different parameter choices. We note that due to computational
cost, it was not possible to test a significantly reduced value for
$M_{\rm cell}^{\rm max}\left(r_{\rm Jet}\right)$ and so the
simulations we present provide the balance between resolution and
numerical feasibility. 

\label{lastpage}
\end{document}